\documentclass{aa}  
\usepackage{graphicx}
\usepackage{xspace}
\usepackage{tabularx}
\usepackage{booktabs}
\usepackage{subfig}
\usepackage{float}
\usepackage{xcolor}
\usepackage{tikz}
\usepackage{float}
\usepackage{xcolor}
\usepackage[normalem]{ulem}
\usepackage{txfonts}
\newcommand{\kms}{$\rm km \, s^{-1}$\xspace}
\newcommand{\cco}{$\rm C^\mathrm{18}O $\xspace(1-0)\xspace}
\newcommand{\co}{$\rm ^{13}CO $\xspace (1-0)\xspace}

\begin{document}

   \title{The Snake Filament: A study of polarization and kinematics }

   \author{Farideh S. Tabatabaei
          \inst{1}
          \and
          Elena Redaelli\inst{2,1}
          \and
          Gabriel A. P. Franco\inst{3}
           \and
          Paola Caselli\inst{1}
          \and
          Marta Obolentseva\inst{1}
           }

   \institute{Centre for Astrochemical Studies, Max-Planck-Institut für extraterrestrische Physik, Gießenbachstraße1, 85748 Garching bei München, Germany
   \and
   European Southern Observatory, Karl-Schwarzschild-Straße 2, 85748 Garching bei München, Germany
    \and
   Departamento de Física—ICEx—UFMG, Caixa Postal 702, 30.123-970 Belo Horizonte, Brazil
        %      \email{farideht@mpe.mpg.de }
             }

   \date{Received XXX; accepted XXX}

% \abstract{}{}{}{}{} 
% 5 {} token are mandatory

  \abstract
 {The role of magnetic fields in the formation of dense filamentary structures in molecular clouds is critical for understanding the star formation process. The Snake filament in or close to the Pipe Nebula's neighboring, a prominent example of such structures, offers an ideal environment to study the interplay between magnetic fields and gas dynamics in the early stages of star formation.  }
   {This study aims to investigate how magnetic fields influence the structure and dynamics of the Snake filament, using both polarization data and molecular line observations. Our goal is to understand the role of magnetic fields in shaping the filamentary structure and explore the kinematics within the filament. }
   { We conducted polarization observations in the optical and near-infrared bands using the 1.6 m and 60 cm telescopes at the Observatório do Pico dos Dias/Laboratório Nacional de Astrofísica (OPD/LNA). Molecular line observations of the \co\ and \cco\ lines were obtained using the IRAM 30m telescope. We analyzed the data to characterize polarization and gas properties within the filament, with a focus on understanding the magnetic field orientation and its relationship with the filament’s structure.}
   { Our findings reveal that the polarization vectors align with the filament’s spine, indicating a magnetic field structure that is predominantly parallel to the filament at lower-density regions. A velocity gradient along the filament is observed in both \co\ and \cco\ lines, with \cco\ tracing the denser regions of the gas. The polarization efficiency decreases with increasing visual extinction, consistent with reduced grain alignment in higher-density regions. The filament’s mass-to-length ratio is below the critical value required for gravitational collapse, indicating stability.}  
   {}

   \keywords{ Stars: formation --
                Magnetic fields --
                Astrochemistry -- 
                ISM: Kinematics and dynamics
               }

   \maketitle
%
%-------------------------------------------------------------------

\section{Introduction}

The study of the formation and evolution of dense filamentary structures in the interstellar medium (ISM) is critical for understanding star formation (\citealt{andre2014}). All molecular clouds contain filamentary structures, as demonstrated in the \emph{Herschel} Gould Belt Survey (\citealt{andre2010}). Furthermore, the majority of gravitationally unstable cores, known as prestellar cores (\citealt{ward2007}), which are on the brink of star formation, are located within these filaments (\citealt{andre2010}, \citealt{konyv2015}). This suggests that dense cores form within the filamentary structures of molecular clouds. Magnetic fields are widely recognized as playing a key role in the formation of these filaments, the cores within those filaments, and the eventual stars within them \citep{Pattle2023}. 
The relative contributions of magnetic fields, turbulence, and gravity to filamentary structure formation remain poorly understood (\citealt{Li2014}, \citealt{Crutcher2012}). Gravity and turbulence influence both the structure and strength of the magnetic field (\citealt{Hennebelle2019}), which in turn shapes the gas dynamics within filaments. Magnetic fields are thought to help organize material into filamentary structures, but their precise role in this process remains unclear, especially in quiescent filaments that may not always channel material toward denser regions.

Dust polarization observations, notably from the Planck satellite, have revealed ordered magnetic field structures in molecular clouds, particularly in the Gould Belt clouds of the Solar neighborhood (\citealt{planck2016}). These observations show that in regions of low column density, the gas structure tends to align parallel to the magnetic field. In contrast, the gas is often oriented perpendicular to the magnetic field in higher-density regions. This alignment transition, observed at visual extinctions of $A_\mathrm{V} \sim$ 2.7 – 3.5 mag, underscores the role of magnetic fields in shaping filamentary structures in star-forming regions (\citealt{planck2016}, \citealt{soler2017}). %Similar behaviors are observed in filaments at lower column densities using optical and near-infrared (NIR) polarization data (\citealt{sugitani2011}, \citealt{palmeirim2013}, \citealt{franco2015}), indicating that magnetic fields play a significant role in the alignment of dust grains within these regions (\citealt{soler2019}, \citealt{santos2016}).
Recent HAWC+ observations of the Serpens South cloud reveal a second transition in the alignment of gas structures with magnetic fields at $A_\mathrm{V} \sim$ 21 mag, where the alignment shifts back to being parallel (\citealt{Pillai2020}). 
%This transition indicates the onset of magnetic supercriticality, allowing gravitational collapse and star cluster formation even in the presence of strong magnetic fields.

In this study, we focus on the Snake filament, also known as Barnard 72, employing both polarization data and molecular line observations to investigate the role of magnetic fields in shaping its filamentary structure. This cloud has received limited attention in the literature; to our knowledge, the only reference to it is found in \citet{Nielbock2012}, where it is mentioned but not studied in detail. Molecular line observations were performed using the IRAM telescope, where we observed the J = 1 --> 0 transitions of $\mathrm{C}^\mathrm{18}\mathrm{O}$ and $^\mathrm{13}\mathrm{C}\mathrm{O}$. Additionally, we obtained polarimetric observations from the 1.6 m and 60 cm telescopes at the Observat\'orio do Pico dos Dias, spanning optical and near-infrared wavelengths. Through the analysis of polarization properties and gas behavior within the Snake filament, we aim to understand how magnetic fields influence its structure and dynamics, thereby advancing our knowledge of star formation in filamentary regions.

This paper is structured as follows: in Sect. \ref{observation}, we detail the observational data obtained from multiple telescopes, including submillimeter polarimetry from Planck, molecular line observations using the IRAM 30m telescope, and optical and NIR polarimetric data from the OPD/LNA telescopes. The methods for data reduction and calibration are also described. In Sect. \ref{results}, we present the analysis and results of data processing focusing on the filament’s polarimetric and kinematic properties, and examining the relationship between the magnetic field geometry and the filament's structure. Section \ref{disscu} investigates the stability of the filament and the polarization efficiency. Finally, in Sect. \ref{conclu}, we summarize the key findings of this work.

%--------------------------------------------------------------------
\section{Data} \label{observation}

%--------------------------------------------------------------------
\begin{figure}[t]
    \centering
    \includegraphics[width=1\linewidth]{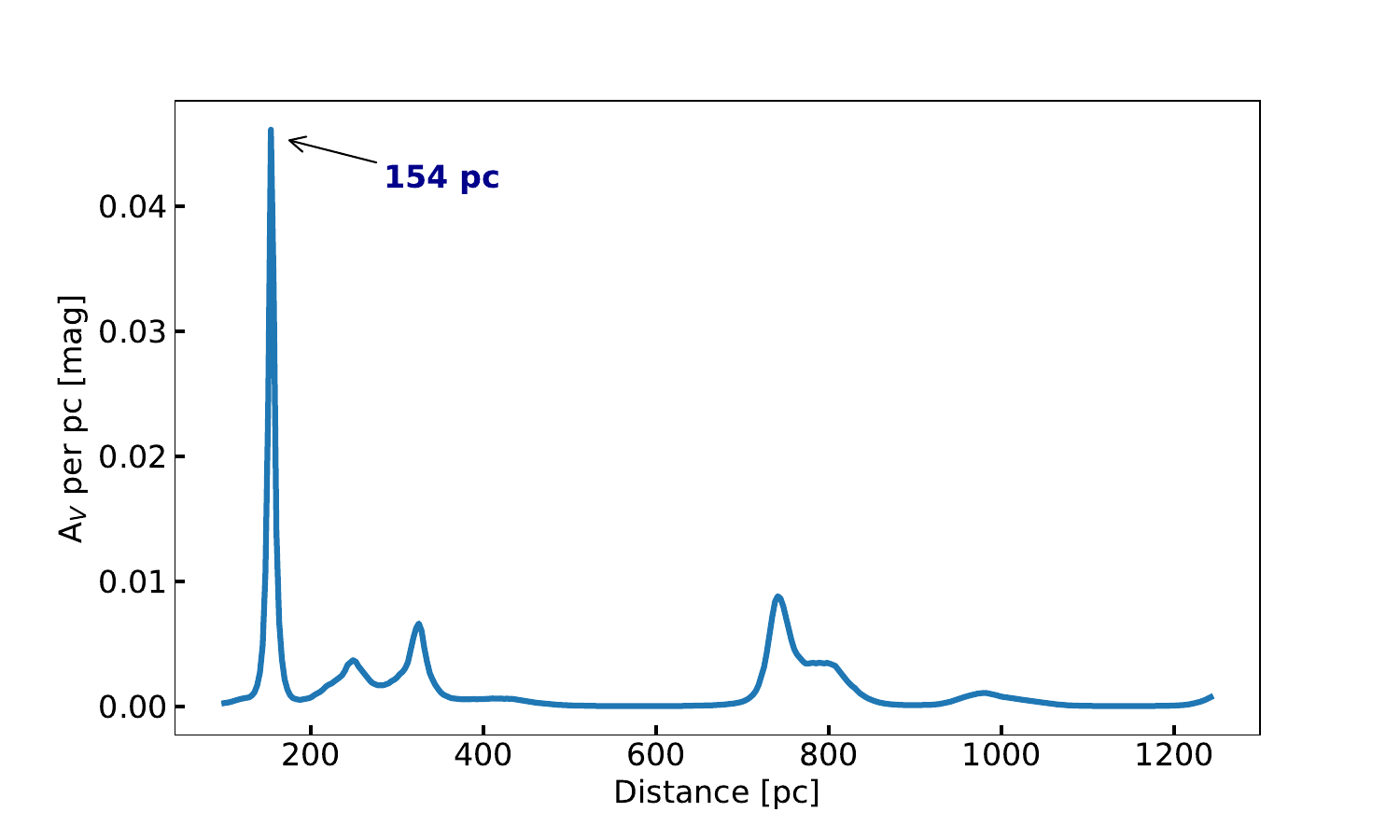}
    \caption{2D dust extinction map illustrating the dust distribution as a function of distance up to 1.25 kpc from the Sun (\citealt{edenhofer2024}). The peak at 154 pc indicates a distance corresponding to the Snake region.}
    \label{distance}
\end{figure}

\subsection{Molecular line data}

The Snake filament was observed using the 30m telescope at Pico Veleta (Spain) from IRAM, under project 038-11, in August of 2011. 
The ground-state transition of \cco and \co at a rest-frame frequency of 110.201354 MHz and 109.782173 MHz was observed, respectively. The observations were made using EMIR E090 receiver, the VErsatile SPectrometer Assembly (VESPA), with a resolution of 20 kHz. The data were calibrated and reduced using CLASS software, GILDAS \footnotetext[1]{https://www.iram.fr/IRAMFR/GILDAS/}\footnotemark[1] software. The mean rms noise level in $T_\mathrm{mb}$ scale is 0.3 K with 0.05 \kms channels.

\subsection{Starlight polarization data}

The polarimetric observations were carried out using the 1.6 m and the IAG 60 cm telescopes at the Pico dos Dias Observatory (OPD\footnotemark[2]/LNA, Brazil)  \footnotetext[2]{The Pico dos Dias Observatory is operated by the Brazilian National Laboratory for Astrophysics (LNA), a research institute of the Ministry of Science, Technology and Innovation (MCTI).} during several missions, 2007, 2008, and 2024 (IAG 60 cm), for the optical data, conducted in the R band (6474 \AA) and, 2018 and 2019, for the near-infrared data (NIR), conducted in the H band. The data were obtained using IAGPOL, an imaging polarimeter specifically adapted for polarimetric measurements, consisting of a rotating retarder $\lambda/2$-waveplate, a calcite Savart prism, and a filter wheel. For further details on the setup of the polarimeter, refer to \cite{maglh96}. 
The optical polarimetry was obtained for 20 fields, with a field of view of about 10$^\prime \times 10 ^\prime$ each, covering the entire region of the Snake filament.
The NIR data were obtained for 11 fields, with a field of view of about 4$^\prime \times 4^\prime$ each. For each optical field, a set of eight positions of the $\lambda/2$-waveplate, positioned at intervals of $22\fdg5$, was obtained with exposure of 100 to 120 s at each position. In 2024 we used the IAG 60 cm telescope to obtain additional optical data. For the NIR, 60 images of 10 s each were acquired for each position of the $\lambda/2$-waveplate, while the telescope was dithering to remove the thermal background signal, adding up to a total exposure of 600 s per $\lambda/2$-waveplate position. The same procedure was repeated for each of the 11 fields of view.

Image processing was performed using the reduction pipeline SOLVEPOL (\citealt{ramirez2017}). After standard procedures of image reduction, the pipeline calculates the differential aperture photometry of the two components generated by the calcite prism, at each position of the wave plate. Observations of standard stars of known polarization angle were used to determine the correction of the polarization angle to the equatorial coordinate system, according to which the position angle is measured East from the celestial north pole. Unpolarized standard stars were also observed in order to determine the instrumental polarization, which turned out to be smaller than 0.1\%. The transformation of pixel to sky coordinates was performed by matching the pixel of one of the two components of each star found in the reduced images with celestial coordinates of their counterparts in the Gaia DR3 catalog \citep{Gaia_Collaboration2023}. The final rms of the transformation solutions was about $0\farcs 1$.

In this study, we assumed that the angle of optical and NIR starlight polarization, denoted as $\phi_\mathrm{star}$, is identical to the orientation of the magnetic field, $\psi$.

%--------------------------------------------------------------------

\subsection{Polarized dust emission from Planck}

We used the 353 GHz Planck\footnotetext[3]{http://pla.esac.esa.int/pla/}\footnotemark[3] polarization data at a resolution of 4$^\prime$.8. From the Stokes Q and U, we study the polarization angles of dust emission in the sky region containing the Snake filament. To improve the signal-to-noise ratio (S/N) of the extended emission, we applied a Gaussian convolution to the Planck beam, achieving 10$^\prime$ resolution maps. We derived the polarization angle of dust emission\footnotetext[4]{We downloaded the data from https://pla.esac.esa.int/, which uses the IAU convention. If the data are downloaded from https://irsa.ipac.caltech.edu/applications/planck/, the derived polarization position angle should be $\phi = 0.5 \: \mathrm{tan}^{-1} (-U/Q)$.}\footnotemark[4], $\phi = 0.5 \: \mathrm{tan}^{-1} (U/Q)$.
The polarized intensity is computed as:
\begin{equation}
P = \sqrt{Q^2 + U^2}. \quad
%\sigma_P = \sqrt{ C_{QQ}^2 +  C_{UU}^2}.
\end{equation}
%where  $C_{QQ}^2$ and $C_{UU}^2$ are the diagonal terms of the covariance matrix.
The Stokes parameters obtained from the Planck database are measured eastward from the north Galactic pole. These measurements were then converted to the equatorial coordinate system, with angles measured eastward from the north celestial pole. We applied the method suggested by \cite{corradi}. Using the equation:
\begin{equation}
\text{offset angle} = \arctan\left( \frac{ \cos(l - 32.9^\circ) }{ \cos b \cot 62.9^\circ - \sin b \sin(l - 32.9^\circ)} \right),
\end{equation}

and considering the Snake filament has galactic coordinate $l \sim 1.8^\circ$ and $b \sim 6.9 ^\circ$, we calculate the offset of $56^\circ$.
Since the area covered by the Snake filament is relatively small.The error introduced by a common offset is smaller than 0.25\textdegree \, in the borders of the studied area. We applied the same offset all over the field, and 56\textdegree \, was subtracted from $\phi$. 

The submillimeter polarization is assumed to be perpendicular to the magnetic field orientation, $\psi$. We rotate the angles by 90\textdegree \, to obtain the corresponding plane-of-sky magnetic field orientations, $\psi_\mathrm{submm}$. Therefore, throughout this work, the polarization data from optical, near-infrared, and submillimeter wavelengths consistently show that the polarization vectors align with the magnetic field direction in their respective regions.
%--------------------------------------------------------------------

\subsection{Data handling and cross-match with \texttt{StarHorse} catalog} \label{starhorse}

For our analysis throughout this paper, we utilize data that correspond to entries in the \texttt{StarHorse} catalog (\citealt{starhorse}). The \texttt{StarHorse} catalog provides stellar parameters, e.g., distances and extinctions, for 362 million stars, by combining Gaia EDR3 data \citep{Gaia_Collaboration2016, Gaia_Collaboration2021} with photometric surveys such as Pan-STARRS1 (\citealt{chambers2016}), SkyMapper (\citealt{Onken2019}), 2MASS (\citealt{Cutri2003}), and AllWISE (\citealt{cutri2013}). Its improved precision and broad wavelength coverage enable distance accuracies of approximately 3\% for stars with a magnitude of $14$ and 15\% for magnitudes of $17$ (\citealt{starhorse}). 
As we discussed in Sect. 3, the Snake Nebula appears to be the dominant structure toward the studied line of sight, up to a distance of 1.2 kpc; in addition, at that distance, the line of sight is at a height of $\sim$150 pc above the Galactic plane, avoiding the densest regions of the Galactic disk, as indicated by the lack of significant cloud material beyond 1 kpc in \citet{vergely2022}. Because of that, we restricted our analysis to stars located within 2 kpc. This selection minimizes uncertainties associated with both distance estimates and polarization measurements, which tend to increase for more distant and fainter stars.
Consequently, all analyses and visualizations presented herein focus solely on stars within a distance of 2 kpc.
In the NIR dataset, we identified 1244 stars, with 491 of these entries corresponding to the \texttt{StarHorse} catalog. Notably, 57 stars are situated within a distance of less than 2 kpc from the Sun. Conversely, the optical dataset includes 19627 stars, of which 13252 match entries in the \texttt{StarHorse} catalog. Among these, 2037 stars are located within 2 kpc of the Sun.

\subsection{Archive data}

We employed archival data from the Gould Belt Survey, collected with the \emph{Herschel} Space Observatory, to derive the dust temperature map (\citealt{andre2010, roy14}).
A visual extinction map was constructed from a deep near-infrared (NIR) imaging survey, combining data from the New Technology Telescope (NTT), the Very Large Telescope (VLT), the 3.5-meter telescope at the Centro Astronómico Hispano Alemán (CAHA), and the 2MASS survey.
(for more details see \citealt{roman10}).

\section{Analysis and Results } \label{results}

\subsection{Distance of the cloud and trends with stellar distance } \label{Sec:distance}

We used the 3D dust extinction map provided by \cite{edenhofer2024} to obtain the dust distribution within the Snake filament region. This map, which extends up to 1.25 kpc from the Sun with a 14$^\prime$ angular resolution, is based on data from 54 million stars as analyzed by \cite{zhang2023}. They derived the stars' atmospheric parameters, distances, and extinctions by forward-modeling the low-resolution Gaia BP/RP spectra (\citealp{carrasco2021}).
We employ the publicly accessible version of the map\footnotetext[1]{https://zenodo.org/records/8187943.}\footnotemark[1].
Figure \ref{distance} shows the dust extinction distribution as a function of distance along the line of sight toward the Snake filament. To obtain a representative profile, we computed the average extinction over a grid of Galactic coordinates covering the main body of the filament ($l = 1.60\textdegree$–$2.09\textdegree$, $b = 6.58\textdegree$–$7.60\textdegree$).  We scaled the map values by 2.8 to convert them to $A_\mathrm{V}$ in magnitudes based on the recommendation of \cite{edenhofer2024}. This averaging reveals a well-defined extinction peak at a distance of 154 pc, which we interpret as the dominant contribution from the Snake filament (Fig. \ref{distance}). This distance agrees with the estimated distance of the entire Pipe Nebula ($145 \pm 16$ pc, \citealt{alves2007}; $163\pm 5$ pc, \citealt{dzib2018}).
%--------------------------------------------------------------------

%

%------------------------------------------------
\begin{figure}[t] % 't' ensures the figure is placed at the top of the page
\centering
\includegraphics[width=0.48\textwidth]{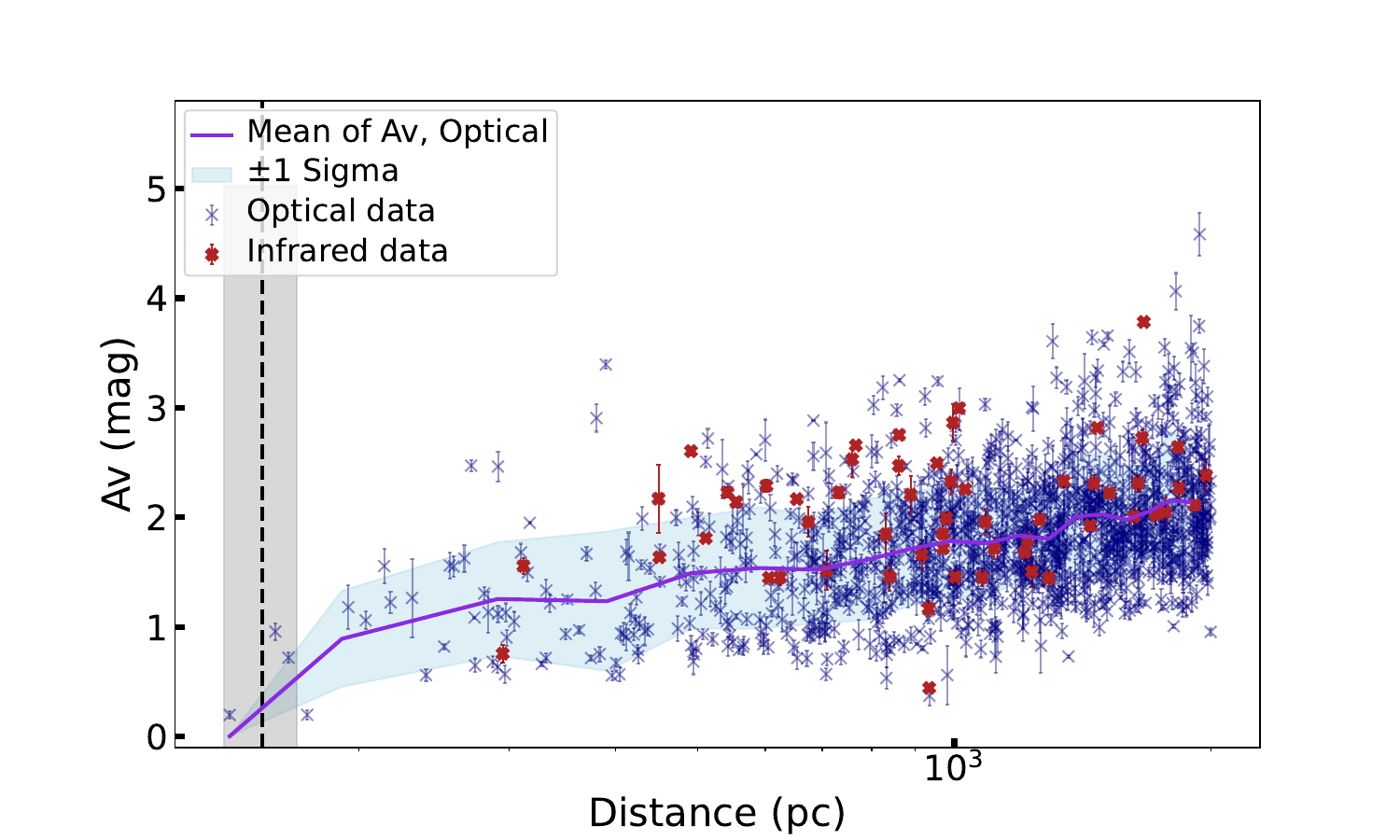}
\includegraphics[width=0.48\textwidth]{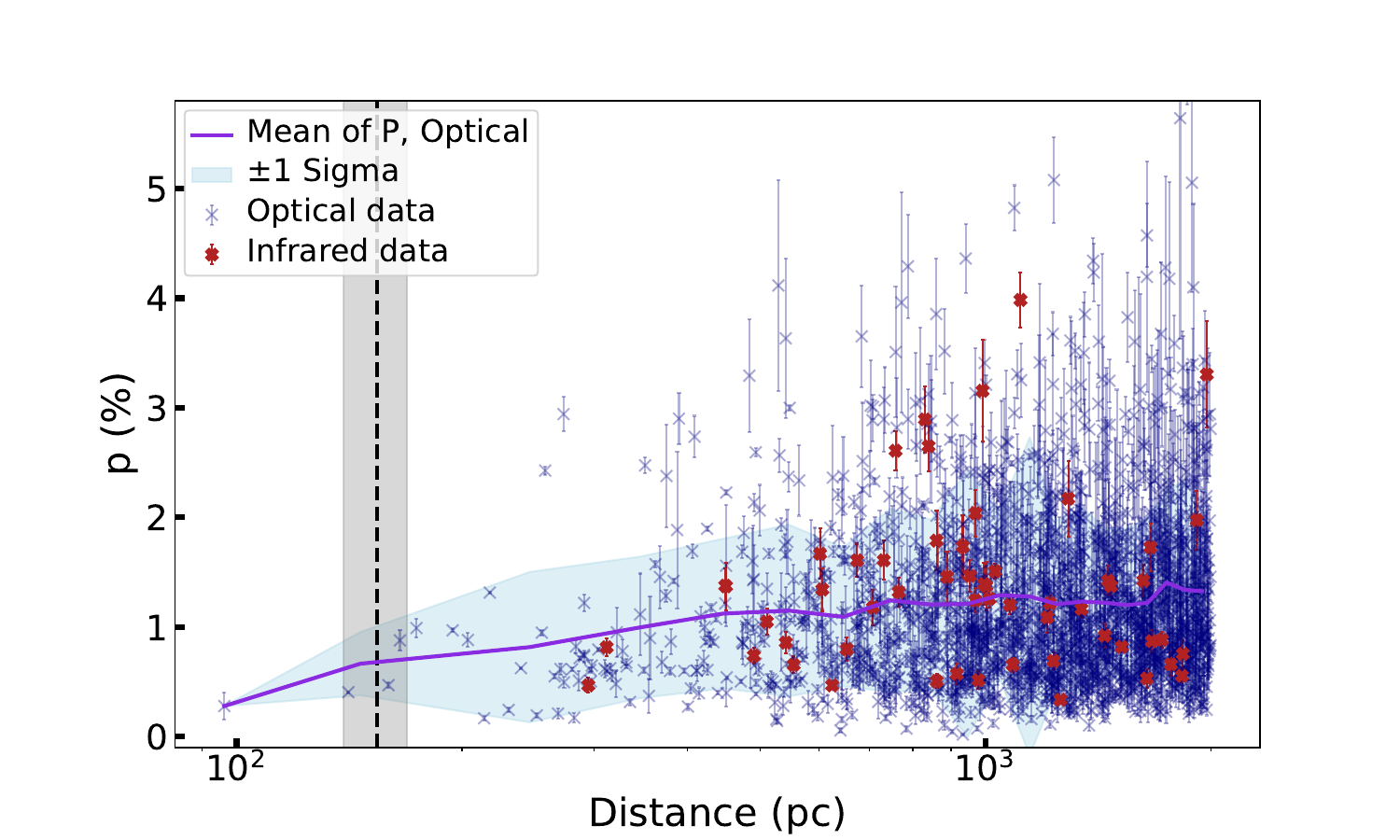}
\caption{Top panel: Visual extinction ($A_\mathrm{V}$) as a function of the distance of stars within 2 kpc from the Sun. The extinction and distance values are derived from the \texttt{StarHorse} catalog. Lower panel: Polarization percentage as a function of the distance. A vertical dashed line at $154\pm 15$ pc marks the distance of the Snake filament. The error bars represent the uncertainties in the extinction and polarization measurements, scaled by 50\%. } The purple line represents the mean value in bins of 100 pc, with the light blue shaded area indicating the \(\pm 1 \sigma\) uncertainty.
\label{Dist}
\end{figure}

Figure \ref{Dist} (upper panel) illustrates the trend of increasing visual extinction as a function of distance, characterized by a gradual slope for stars with distances up to 2 kpc. The blue crosses represent data points from the optical dataset, while the red crosses correspond to the NIR dataset.
A vertical dashed black line at \(d = 154\) pc marks the distance of the Snake filament with the shaded gray area surrounding it (spanning from 139 pc to 169 pc) representing the region of uncertainty, defined as \(\pm 15\) pc. The plot also includes the mean visual extinction (\(A_\mathrm{V}\)) calculated in bins of 100 pc, shown with a purple curve. This line offers a smoothed representation of the overall trend of increasing extinction with distance, helping to highlight the general pattern across the dataset. The light blue area surrounding the purple line shows the \(\pm 1 \sigma\) variation from the mean, providing an uncertainty in the extinction measurements over distance. The optical and NIR datasets show similar trends, while fewer data points are in the NIR dataset.
The bottom panel of Fig.~\ref{Dist} presents the polarization percentage as a function of distance using the same dataset. The mean polarization percentage, calculated in 100 pc bins, reveals an initial increase in polarization, followed by a relatively steady trend extending up to 2 kpc. This steady trend indicates the absence of any dense clouds located beyond the Snake filament, up to the 2 kpc threshold. 
For Fig. \ref{Dist}, we used S/N > 1 for the optical data to include low-polarization stars, which are likely located in the foreground of the Snake filament. For NIR data, we applied S/N > 3, as the observing process for NIR is inherently noisier, resulting in lower S/N values.
For the remaining analysis, we used a more stringent criterion of S/N $\ge$ 5 for the optical data and S/N $\ge$ 3 for the NIR data, to ensure data reliability while retaining sufficient sources.

%--------------------------------------------------------------------
\begin{figure}[t]
   \centering
   \includegraphics[width=\hsize]{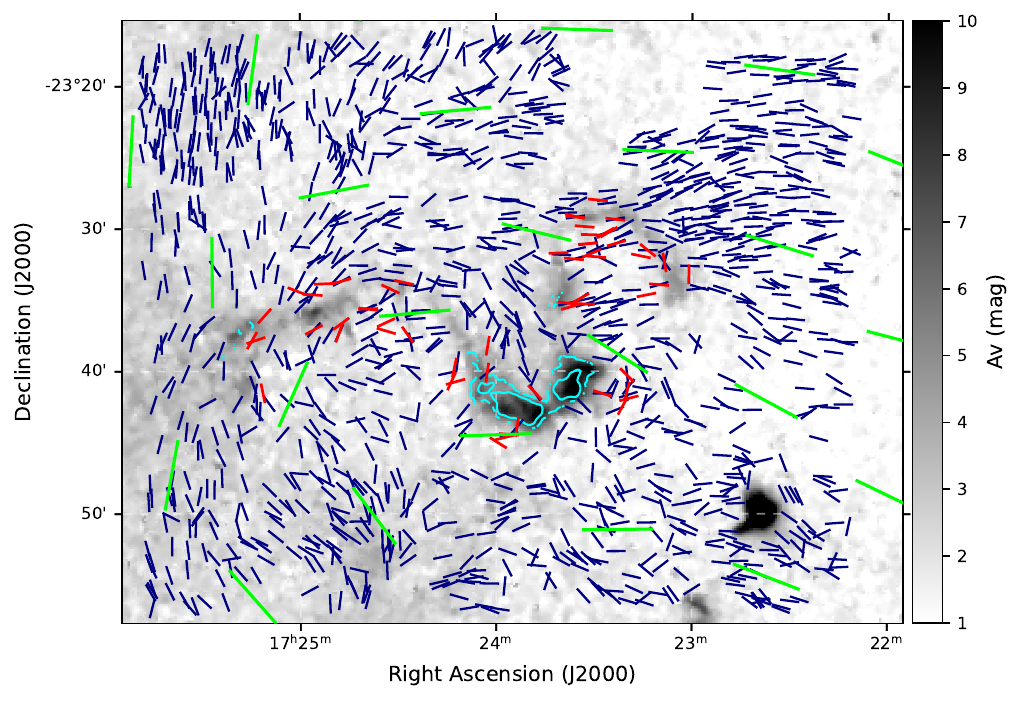}
      \caption{Dust extinction map of the Snake region in color scale at a spatial resolution of 20$^{\prime \prime}$ from \cite{roman10}. All segments represent the B-field orientation. Blue and red segments are inferred from the optical and NIR data, respectively, with S/N$\geq$3. The green segments show the magnetic field orientation from $Planck$ data at the resolution of 10$^\prime$. The vectors are scaled to the same size to enhance visualization clarity. The cyan contours represent the \cco\ integrated intensity at levels of 0.8 K km/s and 1.6 K km/s.} 
         \label{pol_all}
\end{figure}
%%--------------------------------------------------------------------

\subsection{Spatial distribution of the magnetic field}

Figure \ref{pol_all} shows the dust extinction map of the Snake filament region at a spatial resolution of 20$^{\prime \prime}$ (\citealt{roman10}). Polarization vectors in optical (blue), near-infrared (NIR) (red), and submm (green) are plotted over the extinction map. 
%The red and orange vectors indicate the mean polarization in each field of view (10$^\prime \times 10 ^\prime$) for the NIR and optical data sets, respectively. Thus, we can easily follow the orientation of vectors over the observed field. 
The submm vectors are rotated by 90\textdegree \, to show the B-field orientation. The NIR data probe the denser part of the Snake filament, whereas the optical data trace the lower visual extinction of the observed field.
%The magnetic field orientation of submm from Planck data (green vectors) is in agreement with the orientation of optical and NIR data.

\begin{figure}
\centering

\includegraphics[width=0.48\textwidth]{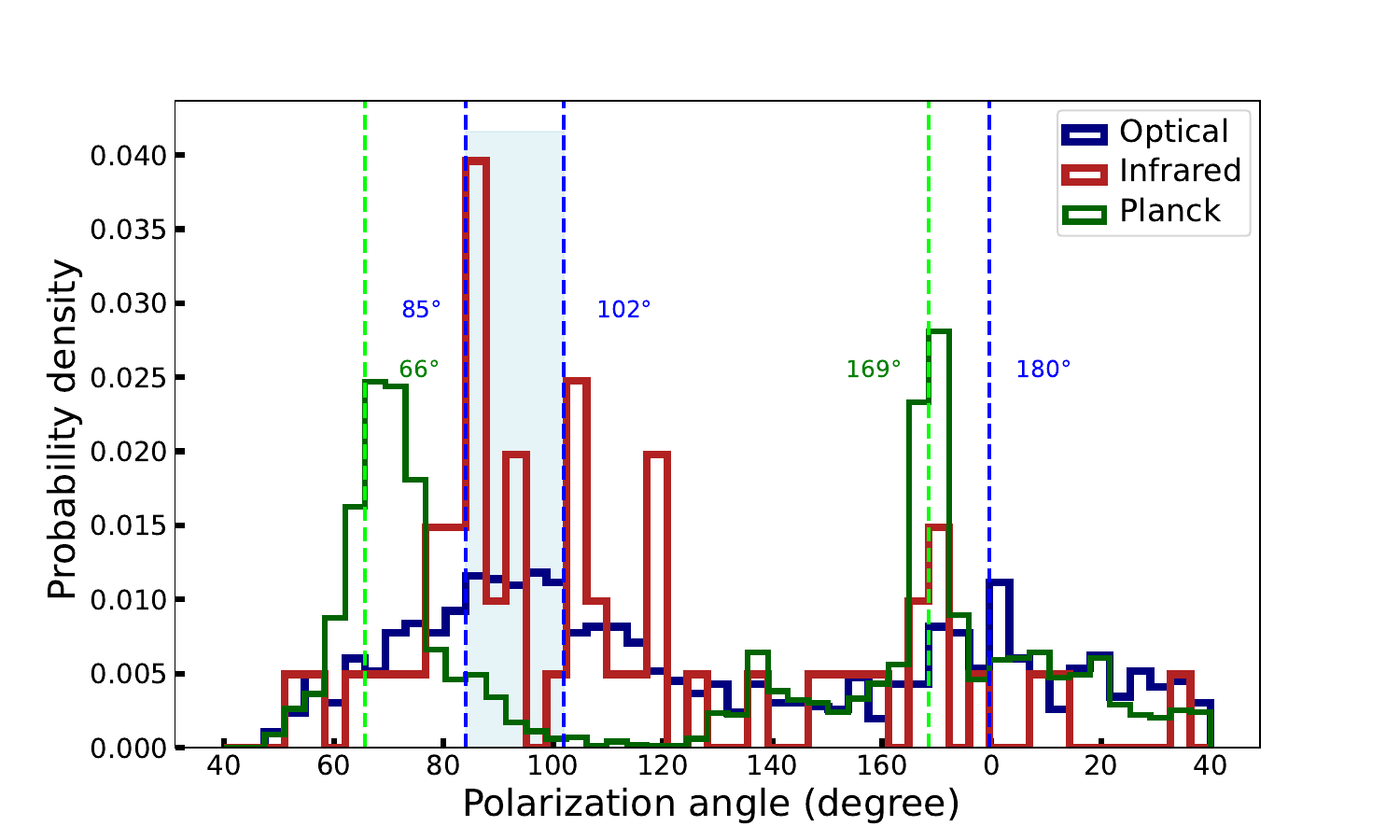}
%\hfill    
        
\caption{ Distribution of polarization angles. Histogram of the distribution of magnetic polarization angles for the three datasets in optical, NIR, and submm. Optical data peaks are highlighted by blue-colored dashed lines, whereas submm dataset distribution peaks are highlighted by green-colored dashed lines.  }
     \label{distri}
\end{figure}

Figure \ref{distri} reveals two prominent peaks in the field lines orientation angles. These two peaks are visible in the optical (blue), NIR (red), and submm (green) data, with their respective values marked as dashed vertical lines in the plot. Specifically, the optical and NIR data exhibit their first peaks within the range of approximately 85\textdegree\, to 102\textdegree, as highlighted in light blue in Fig.~\ref{distri}. The optical data also show a second prominent peak at around 180\textdegree. In contrast, the submm data reveal two distinct peaks at 66\textdegree\, and 169\textdegree, with the second peak of the NIR data aligning closely with the second peak observed in the submm data.
%The accompanying plot (left panel of Fig.\ref{distri}) depicts the polarization angles as a function of right ascension, providing additional insights into how these peaks are associated with specific regions in the studied area. This visualization emphasizes the relationship between polarization angle and spatial distribution. As the right ascension decreases, the polarization angle transitions from vertical to horizontal, illustrating the dynamic nature of the polarization orientation (with respect to North) across the observed field.
In contrast, the infrared data (in red color) illustrates a random distribution in Fig. \ref{distri}. The overall polarization distribution across the entire region does not appear to follow any specific pattern. However, it is important to note that the number of reliable NIR polarization detections is limited compared to the optical data, which reduces the statistical significance of any observed trends in the angle distribution. Despite this limitation, a peak in the distribution can be observed around $80^\circ$ to $100^\circ$.

%--------------------------------------------------------------------
%--------------------------------------------------------------------

\subsection{The relative orientation of the filament with the magnetic field} \label{filametary}

%--------------------------------------------------------------------

 \begin{figure*}
\sidecaption
  \includegraphics[width=12cm]{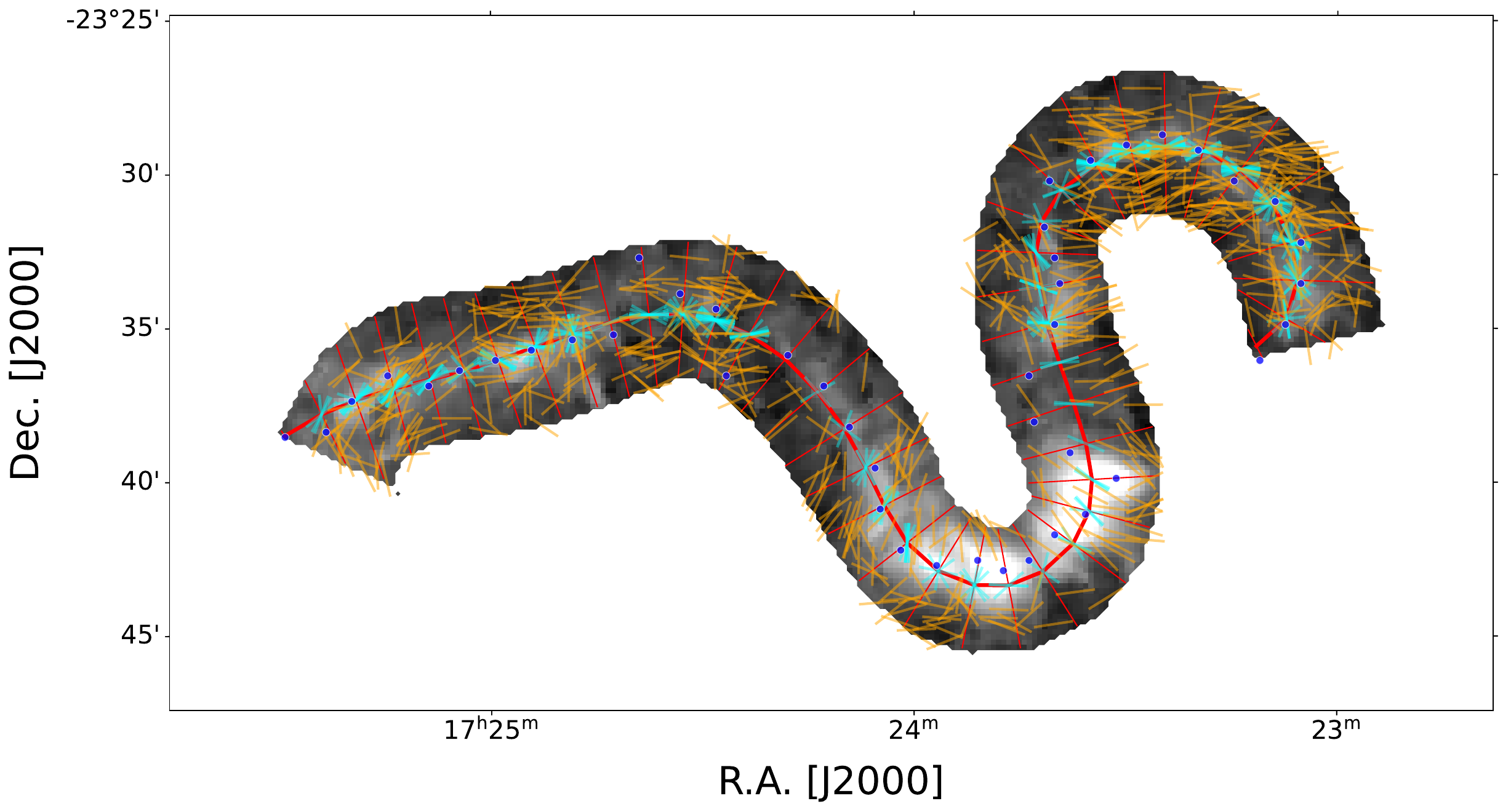}
     \caption{Dust extinction map illustrating the masked filamentary structure. The spine of the filament is outlined by a prominent red curve, while the thin red lines indicate the locations of perpendicular cuts. Blue circles represent peak pixel intensity along each cut. The orange vectors represent optical(S/N$\geq$5) and infrared (S/N$\geq3$) polarization segments. The cyan vectors are projections of all the orange polarization vectors onto the spine. Each polarization vector was projected by finding its closest point on the spine and shifting it accordingly.}
         \label{spine}
\end{figure*}

%--------------------------------------------------------------------
To further characterize the Snake's filamentary structure and its relationship with polarization data, we use the Python package \texttt{radfill} (\citealt{zucker2018}).
The spine of the filament is defined by \texttt{radfil} by using the \texttt{fil\_finder} package (\citealt{koch2015}) based on the Av map and mask map provided. We extracted evenly spaced cuts perpendicular to the spine at intervals corresponding to every 7 pixels along its length. Consequently, the center of the resulting profiles aligns with the peak value of Av. Using this spine, we determine a filament length of 2.3 pc. Figure \ref{spine} illustrates the spine in a thick red curve and the perpendicular cuts. We select a 0.04 degree distance from the spine to make sure we cover the whole snake cloud in our sample (see Fig. \ref{spine}).
The plot also includes polarization data from optical and infrared wavelengths, with the orange vectors representing the polarization angles. The cyan vectors are projections of the orange polarization vectors onto the spine. This projection was done by finding each polarization vector's closest point on the spine and shifting it accordingly.

Our goal is to analyze how the polarization angles change as we move along the filament’s spine. The behavior of polarization vectors along the spine provides critical insights into the magnetic field geometry. Specifically, analyzing the relationship between the polarization vectors and the filament’s curvature allows us to evaluate whether the vectors are predominantly aligned with the filament (parallel) or if there are regions where they deviate significantly (perpendicular).

\begin{figure}
   \centering
   \includegraphics[width=\hsize]{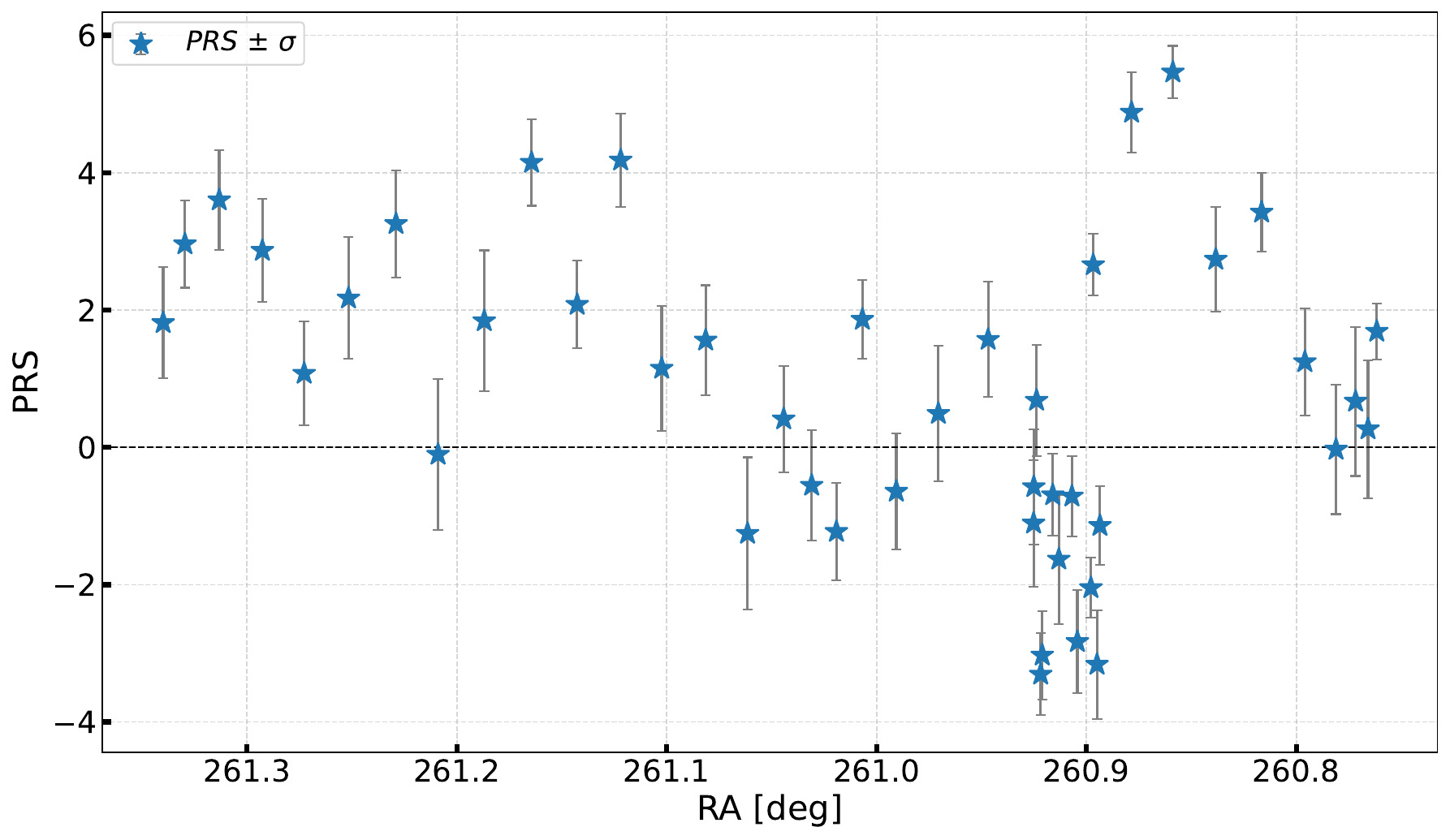}
      \caption{ PRS values as a function of RA. The PRS is calculated at each position along the filament spine using the group of polarization angles projected onto that location. Each point represents the PRS for a given spine position. The dotted line indicates PRS = 0, while the error bars represent the uncertainties.}
         \label{fig:PRS}
   \end{figure}

To quantify the alignment between the magnetic field vectors and the filament orientation, we calculate the Projected Rayleigh Statistic (PRS; \citealt{jow2018}). This method provides a statistical measure of whether the polarization angles are preferentially aligned parallel or perpendicular to the filament's local orientation (tangent to the spine).
The filament spine, shown as the red curve in Fig. \ref{spine}, is divided in 44 positions along its length. For each of these positions, we compute the PRS value using all polarization vectors projected onto that spine location (cyan vectors in Fig. \ref{spine}), determined based on the closest spatial distance.
Following \citet{jow2018}, the PRS is defined as:
\begin{equation}
PRS = \frac{\sum_{i}^{n} \cos \theta_i}{\sqrt{n/2}},
\end{equation}
where $\theta$ is twice the difference between the filament spine angle and the polarization angle. PRS $>>$ 0 indicates strong parallel alignment, and PRS $<<$ 0 indicates strong perpendicular alignment. The uncertainty associated with the PRS value is from (\citealt{jow2018}):
\begin{equation} \label{prs_err}
\sigma^2 = \frac{2\sum_{i}^{n} (\cos \theta_i)^2 -  \mathrm{PRS}^2}{n}.
\end{equation}
Figure \ref{fig:PRS} shows the PRS values as a function of right ascension (RA), with error bars representing the uncertainties at each position along the spine. At both ends of the Snake filament, the polarization vectors are predominantly aligned with the spine, as indicated by PRS values greater than zero, above the dotted line. In contrast, near the central, denser region of the filament, the polarization vectors tend to become more perpendicular to the spine, reflected in the negative PRS values.
The choice of maximum radius from the filament spine affects the PRS values, as it determines which polarization vectors are included in the analysis. To assess this effect, we performed the PRS calculation using a smaller radius of 0.02° (compared to a distance of 0.04° from the spine). The comparison reveals that approximately 9\% of data show a change in the sign of their PRS value (from positive to negative or vice versa). This suggests that while the exact numerical values of PRS may vary slightly with radius, the overall conclusion regarding the alignment of the magnetic field with the filament spine remains robust within this tested range. We are committed to using a radius of 0.04° in our analysis, as this selection ensures that the entire Snake filament is encompassed, as defined by the boundaries of the dust extinction map.

%

%--------------------------------------------------------------------
\subsection{Spectral line fitting}

\begin{figure*}[h!]
\centering
\begin{tikzpicture}
    \node[anchor=north west] at (0,0) {\includegraphics[width=\textwidth]{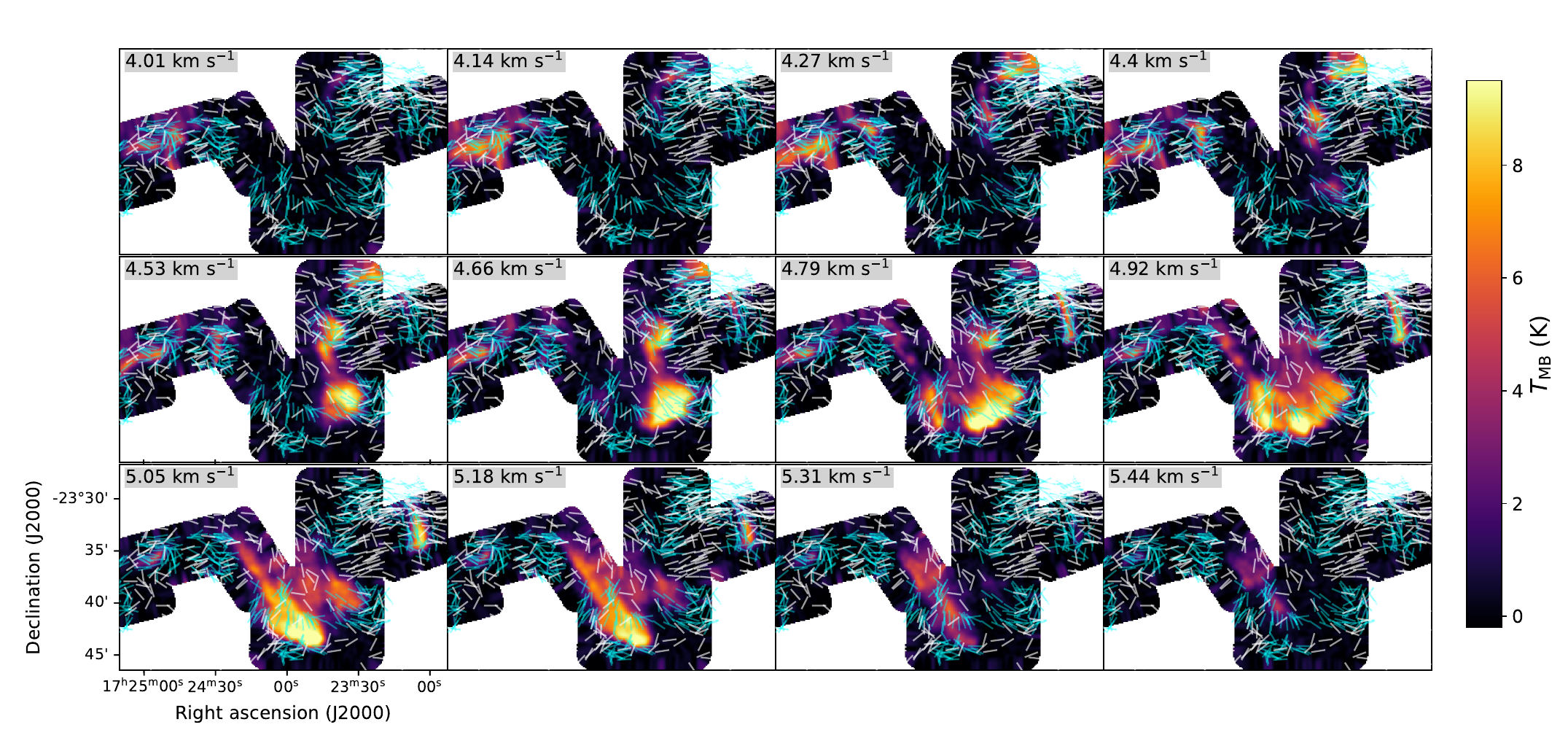}};
    \node[anchor=north west] at (0,-1) {\textbf{a)}}; % Adjust position for "a"
\end{tikzpicture}
\begin{tikzpicture}
    \node[anchor=north west] at (0,0) {\includegraphics[width=0.8\textwidth]{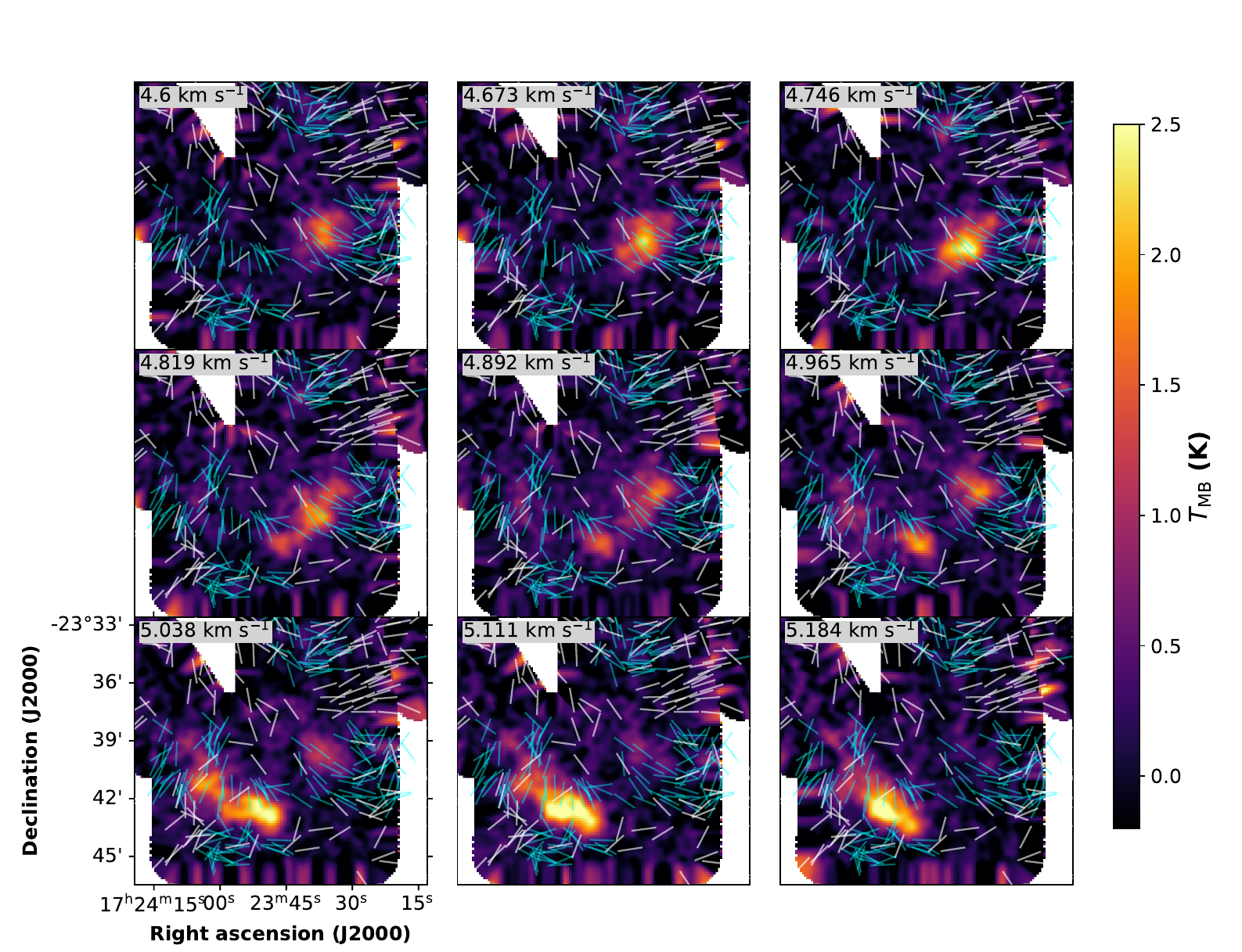}};
    \node[anchor=north west] at (0,-1) {\textbf{b)}}; % Adjust position for "b"
\end{tikzpicture}
\caption{(a) Channel maps between 4.01 and 5.44 \kms\ for the \co\ spectral cube, in 0.13 \kms\ steps. White and cyan segments illustrate B-field orientation in the optical (S/N$\geq$5) and infrared (S/N$\geq$3) bands, respectively. The velocity of each channel is shown on the top left of the panels. (b) Same as (a), but for the \cco\ line. Channels are between 4.60 and 5.18 \kms, in 0.07 \kms\ steps.}
\label{channel_13co}
\end{figure*}

%--------------------------------------------------------------------
%--------------------------------------------------------------------
\begin{figure}
   \centering
   \includegraphics[width=\hsize]{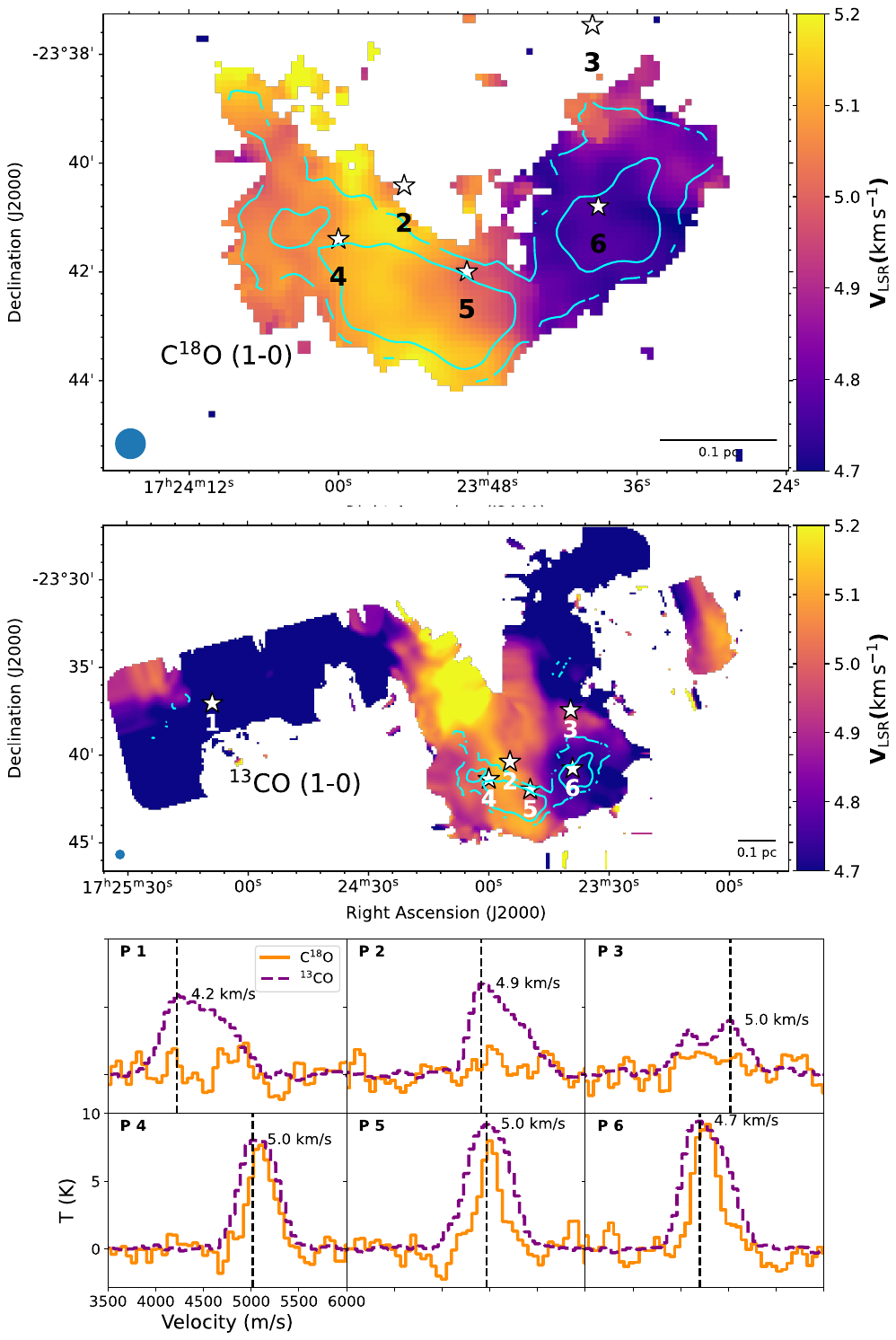}
      \caption{Top panel: Centroid velocity map of the \cco line. Middle panel: Centroid velocity map of the \co\ line. In both maps, we highlight six selected profiles at specific positions along the Snake, marked by white stars. The cyan contours represent the \cco\ integrated intensity at levels of 0.8 K and 1.6 K. The beam size is shown in the bottom-left corner, while the scale bar is displayed in the bottom-right corner. Bottom panel: Comparison of the six spectral profiles corresponding to the positions marked by white stars in both maps. The orange dashed profiles represent the \cco\ line multiplied by 5, while the purple solid profiles correspond to the \co\ line. The black dashed lines indicate the peak velocity of the \co\ line, with the velocity values labeled next to them.}
         \label{velocities}
\end{figure}
%%--------------------------------------------------------------------

CO isotopologues are widely utilized to probe gas in molecular clouds at varying densities. The \co isotopologue is typically associated with tracing the diffuse outer regions of molecular clouds. In contrast, the rarer \cco line, with a critical density of $\sim 10^3 \, \text{cm}^{-3}$, serves as an excellent tracer for regions with higher column and volume densities (relative to \co), effectively avoiding saturation. In this section, we compare the polarization data with the gas kinematics using the complete set of NIR observations rather than limiting the analysis to data cross-matched with the \texttt{StarHorse} catalog (see Sect. \ref{starhorse}). We established that no significant structures exist behind the Snake filament; therefore, all observed NIR data points are considered valid for this analysis.

Figure \ref{channel_13co} displays the \co and \cco channel maps in 0.13 and 0.07 \kms steps, respectively. For both lines, the data presented have an S/N>3. Polarization vectors for optical (white) and NIR (cyan) wavelengths are overplotted on each panel. The \co\ emission at velocities lower than 4.4 \kms\ is confined to the east region of the filament, while higher velocities show more concentrated emission at the center of the filamentary structure. At the filament's center, velocities range from 4.53 \kms\ to 5.18 \kms, transitioning from west to east (see second row).
While the \cco\ emission is only concentrated at the center of the filament, it appears at lower velocities on the western side of the map and shifts to higher velocities toward the east side. 

The kinematic properties of the observed \cco molecular line were derived by fitting Gaussian profiles to each spectrum in the dataset using the Python package \texttt{pyspeckit} \citep{ginsburg11}. A threshold of $\text{S/N} > 5$ was applied, ensuring the fits were constrained to regions with sufficient signal.
A single-Gaussian fit was performed across the unmasked pixels with initial parameter guesses refined iteratively using neighboring pixel values. Fits with residuals exceeding $2 \times \text{RMS}$ or velocity dispersions ($\sigma_v$) broader than 0.35 \kms\ were masked and excluded. The resulting \cco model cube showed only 0.1 \% of the pixels had a bad fit, demonstrating the robustness of the one-Gaussian fitting approach. The top panel of Fig.~\ref{velocities} displays the centroid velocity map of the \cco\ line. The \cco\ line exhibits a narrow linewidth, ranging from $\sim 0.1$ to 0.2 \kms, indicative of quiescent \cco gas dynamics in this region. Given that the sound speed in the region is $c_\mathrm{c}$ = 0.2 \kms (for $T=15$K, from \textit{Herschel} map), these velocity dispersions suggest that the motions are transonic (see Appendix~\ref{dispersion} for dispersion velocity map and sound speed calculation).

In contrast, the \co line exhibits a more complex kinematics, particularly in the central regions of the Snake filament, where two velocity components are observed. Similar to \cco, a single-Gaussian fit was first applied to pixels with $\text{S/N} > 4$, followed by a dual-Gaussian fit for regions with significant residuals or broad line widths. Through visual inspection, we find that lines broader than 0.7 \kms exhibit profiles consistent with two velocity components. Based on this observation, we established a masking criterion that excludes pixels where the residuals exceed $2 \times \text{RMS}$ and velocity dispersions broader than 0.7 \kms\ in the single-Gaussian fit. 
Accurately fitting the \co\ line with two Gaussian components in the central region of the Snake filament is challenging due to the line's high opacity, which exceeds a value of 1 (see Sect. \ref{tau}). The middle panel of Fig. \ref{velocities} presents the centroid velocity map of the \co\ line, derived using a single-Gaussian fitting.  
The central velocities (\(V_{\mathrm{LSR}}\)) of the \co\ line range approximately from $\sim 4.7$\kms to $\sim 5.2$ \kms, increasing from west to east.
% I got these numbers from Casaviewer; I have a screenshot!
%The average uncertainties in \(v_{\mathrm{LSR}}\) are $\sim  6\times 10^{-3}$ \kms in the regions defined in Sec.~\ref{filametary}. 
A similar velocity gradient is observed in the \cco\ line, with velocities increasing from lower to higher values from west to east side. 
The bottom panel of Fig. \ref{velocities} presents six spectral profiles at different random positions along the Snake filament for both \co\ and \cco\ lines. These profiles highlight the saturation of the \co\ line, which prevents an accurate Gaussian fit. The orange dashed profiles represent the \cco\ line, with its temperature scaled by a factor of 5 for better comparison with the \co\ profile. The purple profiles correspond to the \co\ line. For positions 1 to 3 (first row of the plot), the \cco\ line shows no detectable signal, whereas the \co\ line exhibits a strong signal with two distinct velocity components. In positions 4 to 6 (second row of the plot), both lines show clear signals, each with a single velocity component. The peak velocities of the \co\ and \cco\ lines are in good agreement, as indicated by the black vertical dashed lines.

\begin{figure}
   \centering
   \includegraphics[width=\hsize]{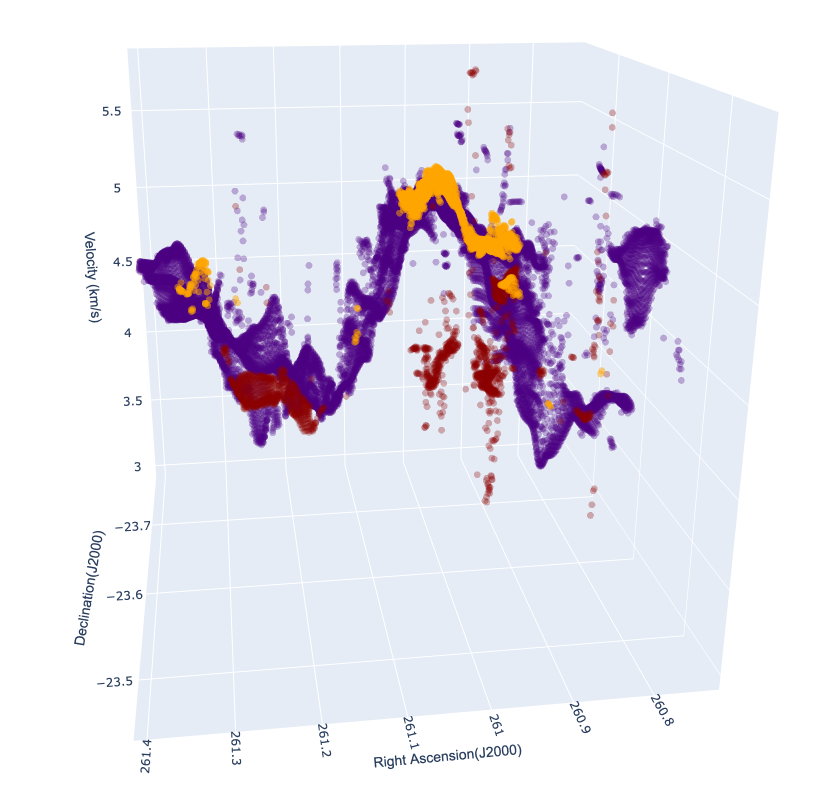}
      \caption{The PPV plot illustrates the kinematics of the \co\ and \cco\ lines. Each data point represents the spatial location and centroid velocity of a Gaussian component, with colors distinguishing different Gaussian fits. The centroid velocity of the \cco\ line is shown in orange, while the purple and red points correspond to the brighter and fainter intensity components of the \co\ line, respectively.}
         \label{ppv}
   \end{figure}

Figure \ref{ppv} presents the 3D position-position-velocity (PPV) diagram, illustrating the \cco\ line in orange and the two velocity components of the \co\ line in purple and red, respectively. The higher velocity component of the \co\ line overlaps significantly with the velocity of the \cco\ line (see the orange data point and purple data point in Fig. \ref{ppv}).
From Figures \ref{velocities} and \ref{ppv}, we conclude that both gas tracers follow the same cloud, exhibiting similar velocities. The gas shows lower velocities at both ends of the Snake filament, with velocities increasing toward the central region of the filamentary structure, which corresponds to the denser part of this region.
This pattern suggests that gas is accreting toward the central region, a behavior also observed in other filaments (e.g., \citealt{feng2024}). As discussed in Sect. \ref{filametary}, the magnetic field lines are aligned with the filament's spine at both ends. We imply that the magnetic field plays a role in guiding material toward the central region of the filament, where the gas density increases. In this denser region, the magnetic field morphology becomes more complex, leading to the observed orientation of magnetic field lines that appear perpendicular to the filament's spine.

%--------------------------------------------------------------------
%
%--------------------------------------------------------------------

%\subsection{Molecular line analysis}
%We use the $\textsc{FilFinder2D}$ package to process the Vlsr map of \cco to identify the spine of the filament (Fig. \ref{velo_c18o}. The spine represents the central axis of the filaments and is crucial for further analysis. 
%We determine the mean velocity for each spine coordinate by averaging the velocities within a 7$\times$7 pixel grid centered on the coordinate. This grid extends 3 pixels in every direction from the central point, accurately representing the velocity associated with the spine. Figure \ref{spine_ra_velo} illustrates the spine\'s velocity behavior in the RA direction. 

%--------------------------------------------------------------------
%\begin{figure}
 %   \centering
  %  \includegraphics[width=1\linewidth]{spine_ra_vs_velocity.pdf}
   % \caption{spine velocity}
    %\label{spine_ra_velo}
%\end{figure}
%--------------------------------------------------------------------

\subsection{Line opacity} \label{tau}

We estimate the opacity of \co and \cco using the following expression:
\begin{equation}
    \tau = - \ln \left[ 1- \frac {T_\mathrm{mb}} {J_\nu(T_\mathrm{ex}) - J_\nu(T_\mathrm{bg})} \right],
\end{equation}
where $T_\mathrm{ex} = 16 $ K is the excitation temperature derived from the \emph{Herschel} dust temperature map, $T_\mathrm{mb}$ is the peak main beam temperature of the line, $J_\nu$ represents the equivalent Rayleigh-Jeans temperature, and $T_\mathrm{bg} = 2.73$ K is the temperature of the cosmic microwave background.
To estimate uncertainties, we varied $T_\mathrm{ex}$ within a physically meaningful range, ensuring that the lower limit does not exceed the minimum excitation temperature, $T_\mathrm{ex} = 15$ K.
The optical depth of \cco\ is determined to be $\tau = 0.6^{+0.1}_{-0.1}$, calculated using $T_\mathrm{mb} = 5.5$ K. 
The uncertainties were estimated by considering a $\pm1$ K variation in $T_\mathrm{ex}$. 
Similarly, for the \co\ line, we obtain $\tau = 2.1^{+0.4}_{-0.9}$ using $T_\mathrm{mb} = 11$ K. 
These results confirm that the \cco\ gas is optically thin, whereas the \co\ gas is optically thick.
The \cco line has a relatively low critical density ($n_c \sim 10^3 \ \mathrm{cm}^{-3}$), making it reasonable to consider that the excitation temperature ($T_\mathrm{ex}$) is approximately equal to the kinetic temperature ($T_\mathrm{k}$), considering dust-gas coupling. Although the \emph{Herschel} dust temperature is typically higher than the excitation temperature, as mentioned earlier, we also calculated the line opacity for a variation in $T_\mathrm{ex}$ as an uncertainty.

\section{Discussion} \label{disscu}

\subsection{Filament stability and absence of star formation}

Filament stability can be assessed by comparing their mass-to-length ratio to a critical value (\citealt{Ostriker1964, fiege2000}). To incorporate the effect of non-thermal motions, we follow the formulation introduced by \cite{hacar2023}:
\begin{equation}
\left(\frac{M}{L}\right)_\mathrm{cr} =\, m_\mathrm{vir}(\sigma_\mathrm{tot}) = \frac{2\sigma_\mathrm{tot}^2}{G} \sim 465 \left(\frac{\sigma_\mathrm{tot} }{1 \, \mathrm{km s^{-1}}}\right)^2 \, \mathrm{M}_{\odot} \, \mathrm{pc^{-1}},
\end{equation}
where \( G \) is the gravitational constant and \( M_{\odot} \) is the Solar mass. $\sigma_\mathrm{tot}$ is the total gas velocity dispersion, $\sigma_\mathrm{tot}^2 = \sigma_\mathrm{nt}^2+c_\mathrm{s}^2$. For detail calculation of the $\sigma_\mathrm{tot}$ see Appendix~\ref{dispersion}.  

%Theoretically, a self-gravitating isothermal cylinder has an infinite radius but a finite mass. The critical mass-to-length ratio at a temperature of 10 K is calculated as \citep{Ostriker1964, fiege2000}: \begin{equation}\left(\frac{M}{L}\right)_\mathrm{cr} = \frac{2c_\mathrm{s}^2}{G} = 16 \left(\frac{T}{10 \, \mathrm{K}}\right) \, \mathrm{M}_{\odot} \, \mathrm{pc^{-1}},\end{equation}where \( G \) is the gravitational constant and \( M_{\odot} \) is the Solar mass. 
If the actual mass-to-length ratio of the filament, \( \left(\frac{M}{L}\right)_\mathrm{fil} \), exceeds the critical value, the filament will collapse radially. To estimate the mass of the filament, we use the following equation:

\begin{equation}
M_\mathrm{fil} = 1.13 \times 10^{-7} \times \sum_{i=1}^n \left(\frac{N(\mathrm{H}_2)}{\mathrm{cm}^{-2}}\right)_i \left(\frac{\theta}{''}\right)^2 \left(\frac{d_\mathrm{cloud}}{\mathrm{pc}}\right)^2 \mu_\mathrm{H_2} m_\mathrm{H} \, M_{\odot},
\end{equation} 
where the summation \( \sum_{i=1}^n (N(\mathrm{H}_2)_i) \) includes contributions from \( n \) pixels within the filament area as introduced in Sect. \ref{filametary}, \( m_\mathrm{H} = 1.67 \times 10^{-24} \, \mathrm{g} \) is the mass of a hydrogen atom, \( \mu_\mathrm{H_2} = 2.8 \) (from \citealt{kauffmann08}) is the mean molecular weight per hydrogen molecule, and \( d_\mathrm{cloud} = 154 \, \mathrm{pc} \) is the distance to the cloud, \( \theta= 10'' \) is the pixel size. 
We calculated the gas column density of the filament, \( N(\mathrm{H_2}) \), from the visual extinction, \( A_\mathrm{V} \) (Fig. \ref{pol_all}). Using the relation provided by \cite{bohlin}, 
$N(\mathrm{H_2}) = 9.4 \times 10^{20} \times A_\mathrm{V} \, \mathrm{cm}^{-2} \, \mathrm{mag}^{-1},$ we converted \( A_\mathrm{V} \) into the molecular hydrogen column density. 
The filament length is estimated to be \( L = 2.4 \, \mathrm{pc} \), with a detailed description of the measurement method provided in Sect. \ref{filametary}. The mass per unit length of the filament is determined to be 14.4 \( M_{\odot} \, \mathrm{pc}^{-1} \), which is below the critical mass per unit length, \( \left(\frac{M}{L}\right)_\mathrm{cr} = 31.3 \, M_{\odot} \, \mathrm{pc}^{-1} \). This indicates that the Snake filament is stable against gravitational collapse. This finding is valid assuming negligible external pressure onto the filament. The magnetic field would only increase the support against collapse. Consequently, this is likely the primary reason for the absence of star formation in this region.
%-----------------------------------------------------------------
%---------------------------------------------------

\begin{figure}
    \centering
    \includegraphics[width=1\linewidth]{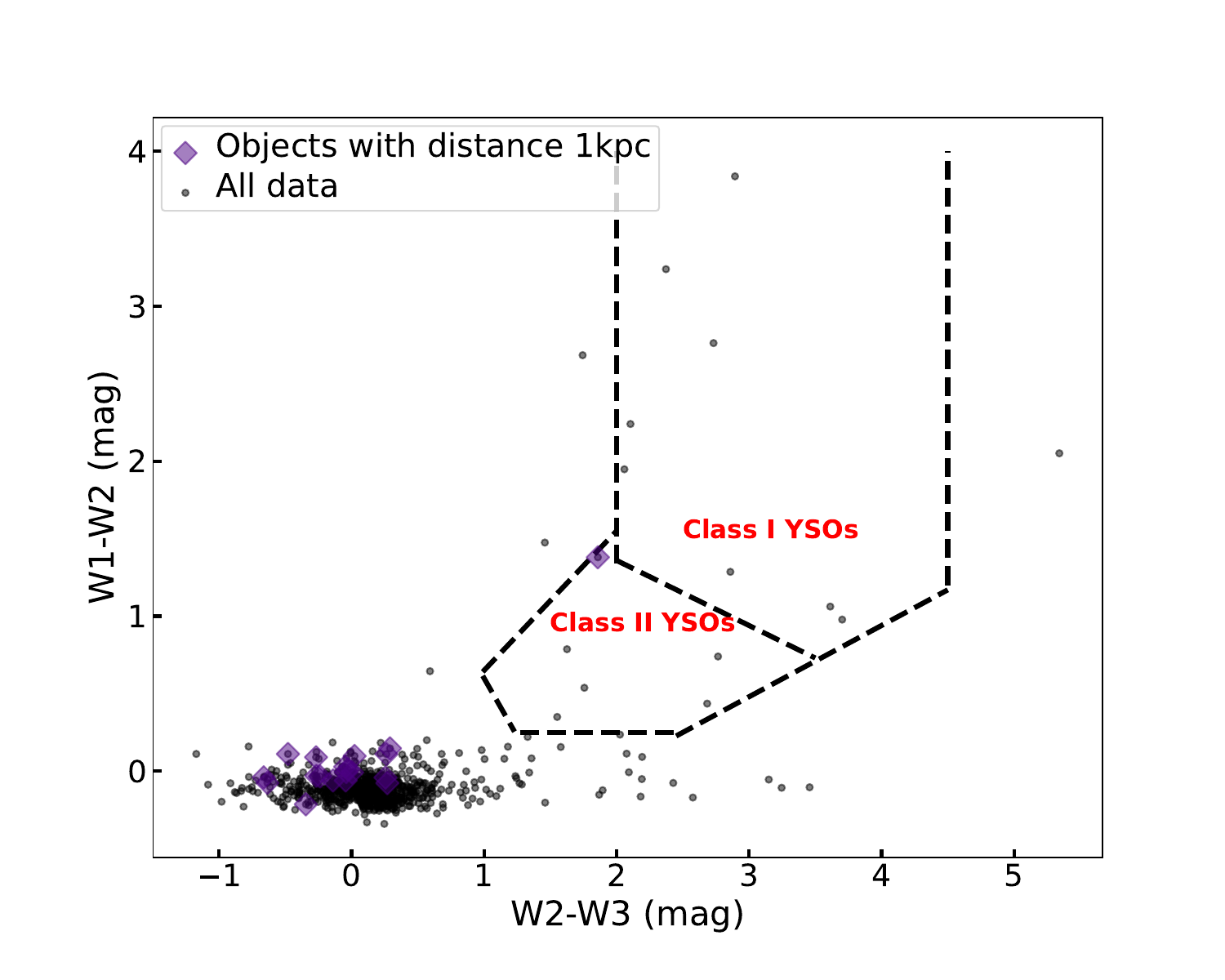}
    \caption{The W1-W2 versus W2-W3 color-color diagram of AGB candidates from the SPLICES survey. The dashed areas used by \cite{koening2014} to identify Class I and Class II YSOs are shown. Black dots represent all the data from SPLICES, and purple diamonds indicate objects within a distance of 1 kpc.}
    \label{yso}
\end{figure}

We checked for the presence of recent star formation activity using version 7.1 of the SPHEREx Target List of ICE Sources (SPLICES), which is part of the SPHEREx all-sky survey mission \citep{ashby2023}. The catalog contains 8.6 \(\times 10^6\) objects that are brighter than approximately 12 Vega magnitudes in the W2 band. To identify young stellar objects (YSOs), we use the W1[3.6 mag] - W2[4.5 mag] versus W2[4.5 mag] - W3[5.8 mag] color-color diagram of AGB candidates from SPLICES \citep[see][]{ashby2023}. The regions defined by \cite{koening2014} for the selection of Class I and Class II young stellar objects.
No YSOs have been detected in the area using the Spitzer GLIMPSE color-color diagram. We utilize the mid-infrared colors [3.6 - 4.5] versus [4.5 - 5.8] to identify YSOs in the region.
We consider a region of size 2 pc that encompasses the filamentary features associated with the Snake region, shown in Fig. \ref{pol_all}, and find 822 objects that are detected in the region (black dots in Fig. \ref{yso}). We apply a distance constraint, selecting objects that are 1 kpc away (represented by purple diamonds in Fig. \ref{yso}), and we only find 19 objects within 1 kpc distance. Among these, we identify only one object as Class II, which is situated far from the Snake region.

%--------------------------------------------------------------------

\subsection{polarization efficiency}
\begin{figure}[h!]
    \centering
    \includegraphics[width=1\linewidth]{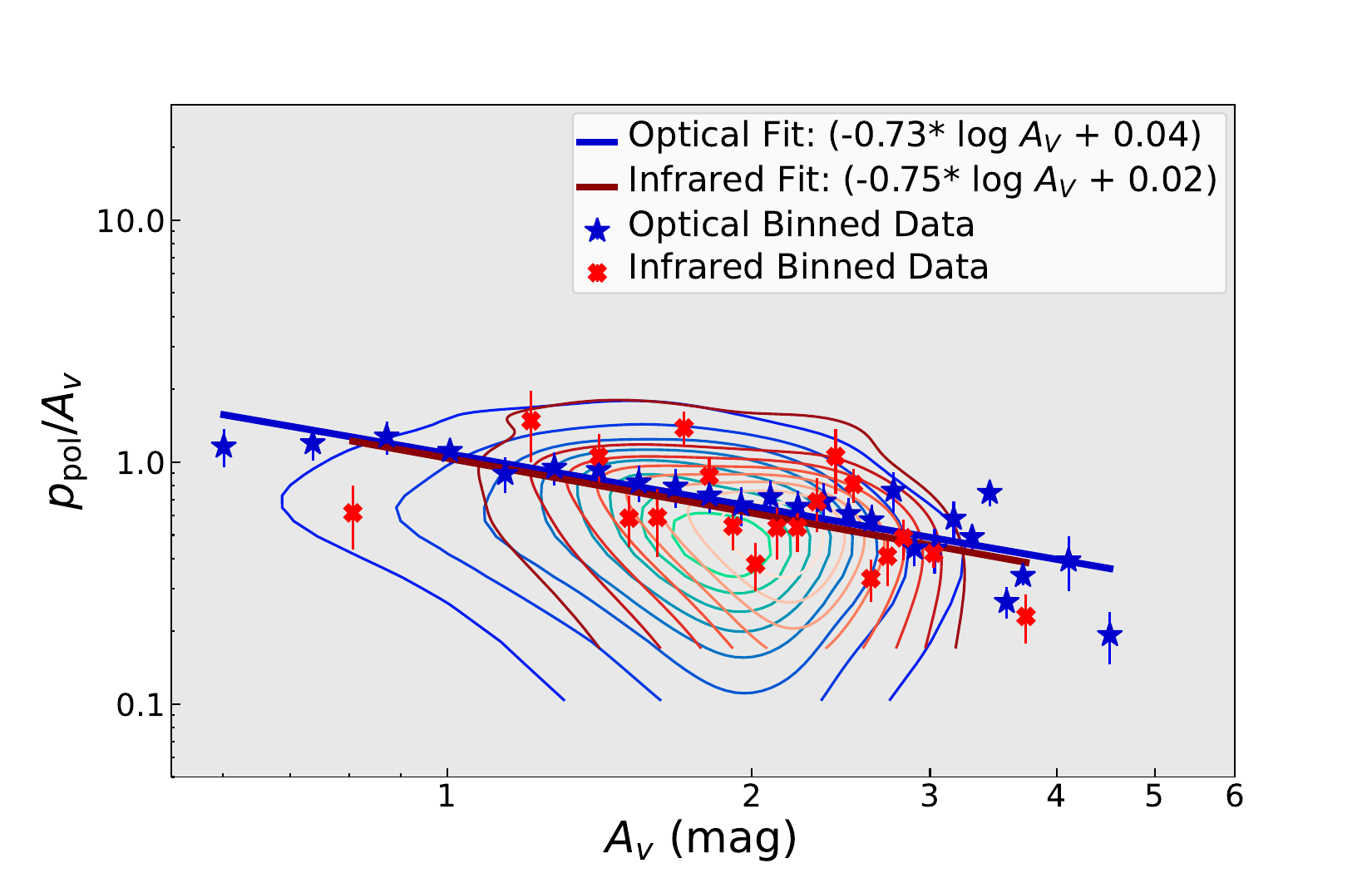}
    \caption{Scatter plots of the polarization efficiency ($p_{\rm pol}/A_{\rm V}$) as a function of the visual extinction ($A_{\rm V}$) in magnitudes in logarithmic scale. Measurement uncertainties are displayed as an error bar for each data point. Solid lines are the best fit for the datasets, as explained in the main text. Contours represent levels of constant probability density estimated using a 2D kernel density estimation (KDE) of the data (using \texttt{seaborn} Python library, \citealt{seaborn}). These contours indicate where the data points are most densely concentrated. Both sets of contours span density levels from 0 to 0.8, with steps of 0.1, effectively highlighting regions of high and low data concentration. The best-fit parameters are shown in the top-right corner. }
    \label{polarisationEfficiency}
\end{figure}

Figure \ref{polarisationEfficiency} presents the scatter plots of polarization efficiency \((p_{\rm pol}/A_{\rm V})\) as a function of visual extinction \((A_{\rm V})\) in log-log space. The optical and NIR datasets are organized into 30 bins, represented by blue and red points, respectively. The error bars on each data point are calculated as the standard deviation of the polarization measurements within each bin, reflecting the measurement uncertainties in the analysis. The blue and red curves illustrate the density plots for the optical and NIR datasets, respectively. Blue and red contours show KDE density levels from 0.0 to 0.8 in steps of 0.1.
The optical and NIR data show a peak density within the visual extinction range of 1 to 3 magnitudes.
%The optical data exhibit a maximum density of 86\% within the range of \(A_V\) from 1 to 3 magnitudes, while the NIR data show a slightly higher maximum density of 46\% within the range of \(A_V\) from 2 to 3 magnitudes.

The plot includes two fitted lines corresponding to two datasets. The optical fit is represented by the equation: $p_{\rm pol}/A_{\rm V} = -(0.73\pm 0.09) \log A_{\rm V} + 0.04$. Conversely, the NIR fit is described by: $p_{\rm pol}/A_{\rm V} = -(0.75\pm 0.27) \log A_{\rm V} + 0.02$, which shows a slightly steeper decline in polarization efficiency with increasing visual extinction, however, both follow the same overall trend, as they span very similar ranges of extinction values.
 The optical polarization is primarily associated with the diffuse cloud surrounding the Snake filament, while the infrared data predominantly trace slightly denser regions. However, both datasets in our analysis correspond to relatively low-density regions within the Snake filament, spanning a visual extinction range of 0.5 to 4.5 magnitudes (see Fig.~\ref{pol_all}). The polarization efficiency exhibits a decrease at both optical and infrared wavelengths. The slopes observed for both datasets indicate depolarization in this region, consistent with the radiative alignment theory (RAT, \citealt{Lazarian2007}). As visual extinction increases, reduced radiation penetration within the cloud diminishes the effectiveness of radiative torques in aligning dust grains. 
 As reported by \cite{alves2014}, NIR polarization measurements of the starless core Pipe-109 in the Pipe Nebula yield $\alpha$ (which is the slope of the fit) values of 1.0 for $A_\mathrm{V} < 9.5$ mag and 0.34 for $A_\mathrm{V} > 9.5$ mag.
\cite{redaelli2019} determined a steep slope of $\alpha = 1.21$ for the FIR polarization data associated with the protostellar core IRAS 15398-3359. Similarly, \cite{tabatabaei2024} derived a slope of 0.75 from NIR data of a filamentary structure in Barnard 59. 
However, other mechanisms may also contribute to this depolarization trend. \cite{Seifried2019} conducted radiative transfer modeling using the \texttt{POLARIS} code and showed that dust grains remain well aligned even at high densities ($n > 10^3$ cm$^{-3}$) and extinctions ($A_V > 1$), provided RAT alignment is active. They found that the frequently observed decrease in polarization degree is primarily caused by large variations in the magnetic field orientation along the line of sight, rather than a breakdown of grain alignment.
\cite{wang2019} reported a power-law slope of $\alpha = 0.56$  from 850 $\mu$m polarization observations of the IC 5146 filament, which implies that dust grains in this $A_V \sim 20$–300 mag range can still be aligned with magnetic fields.
%Overall, the scatter plots demonstrate the relationship between polarization efficiency and visual extinction across optical and infrared datasets, providing insights into the dust properties and the physical conditions of the molecular clouds studied.

%--------------------------------------------------------------------
%--------------------------------------------------------------------

\section{Summary} \label{conclu}

In this work, we investigated the magnetic field geometry and kinematic properties of the Snake filament by combining archival submillimeter polarimetry from Planck and molecular line data from IRAM with new optical and NIR polarization measurements. Our analysis confirms that the Snake is the dominant dust structure along the line of sight within 2 kpc of the Sun, with no significant molecular structures located beyond this distance. As a result, the observed polarization angles of background stars trace the magnetic field associated with the Snake filament. Additionally, we determined the distance of the source to be 154 pc using Gaia data (see Sect. \ref{Sec:distance}). This distance puts the Snake filament at the same distance as the Pipe Nebula.

The Snake filament is gravitationally stable. The estimated mass-to-length ratio is below the critical threshold, explaining the lack of star formation in the region. The polarization efficiency (\(p_{\rm pol}/A_{\rm V}\)) decreases with increasing visual extinction, with fitted slopes of \(0.73\) for optical data and \(0.75\) for NIR data. This behavior suggests depolarization in denser regions where reduced radiation penetration weakens dust grain alignment. 

Kinematic analysis reveals a velocity gradient along the filament in both \co\ and \cco\ data. The \cco\ line, tracing denser regions, exhibits well-constrained single-Gaussian fits, while the \co\ line shows more complex kinematics, including multiple velocity components in the denser central region. Furthermore, the polarization vectors along the filament’s spine exhibit strong alignment with the filament’s spine tangent at both ends. Consequently, it appears that the field may help channel material toward the central region where the gas density increases. In this denser central area, the magnetic field morphology appears more complex, resulting in field lines that are oriented perpendicular to the filament's spine.

These results demonstrate that the Snake filament is immersed in a well-ordered magnetic field, which plays a critical role in maintaining its structural coherence and stability. The observed polarization and kinematics suggest that magnetic fields significantly influence the dynamics of the filament, shaping its evolution. This study contributes to understanding the role of magnetic fields in filamentary structures and their connection to star formation processes.

\begin{acknowledgements}
      Elena Redaelli acknowledges the support from the Minerva Fast Track Program of the Max Planck Society. G. Franco acknowledges the partial support from the Brazilian agencies CNPq and FAPEMIG. We thank the staff of OPD/LNA (Brazil) for their hospitality and invaluable help during our observing runs. This research has used data from the \emph{Herschel} Gould Belt survey (HGBS) project (http://gouldbelt-herschel.cea.fr). The HGBS is a \emph{Herschel} Key Programme jointly carried out by SPIRE Specialist Astronomy Group 3 (SAG 3), scientists of several institutes in the PACS Consortium (CEA Saclay, INAF-IFSI Rome, and INAF-Arcetri, KU Leuven, MPIA Heidelberg), and scientists of the \emph{Herschel} Science Center (HSC). This work has made use of data from the European Space Agency (ESA) mission
{\it Gaia} (\url{https://www.cosmos.esa.int/gaia}), processed by the {\it Gaia}
Data Processing and Analysis Consortium (DPAC,
\url{https://www.cosmos.esa.int/web/gaia/dpac/consortium}). Funding for the DPAC
has been provided by national institutions, in particular the institutions
participating in the {\it Gaia} Multilateral Agreement. The authors express their gratitude to Jaime Pineda for his valuable support with the data reduction process of the IRAM data.
\end{acknowledgements}


\bibliographystyle{aa} % style aa.bst
\bibliography{aanda} % your references Yourfile.bib

\begin{thebibliography}{50}
\expandafter\ifx\csname natexlab\endcsname\relax\def\natexlab#1{#1}\fi

\bibitem[{{Alves} \& {Franco}(2007)}]{alves2007}
{Alves}, F.~O. \& {Franco}, G.~A.~P. 2007, \aap, 470, 597

\bibitem[{{Alves} {et~al.}(2014){Alves}, {Frau}, {Girart}, {Franco}, {Santos}, \& {Wiesemeyer}}]{alves2014}
{Alves}, F.~O., {Frau}, P., {Girart}, J.~M., {et~al.} 2014, \aap, 569, L1

\bibitem[{{Anders} {et~al.}(2022){Anders}, {Khalatyan}, {Queiroz}, {Chiappini}, {Ard{\`e}vol}, {Casamiquela}, {Figueras}, {Jim{\'e}nez-Arranz}, {Jordi}, {Mongui{\'o}}, {Romero-G{\'o}mez}, {Altamirano}, {Antoja}, {Assaad}, {Cantat-Gaudin}, {Castro-Ginard}, {Enke}, {Girardi}, {Guiglion}, {Khan}, {Luri}, {Miglio}, {Minchev}, {Ramos}, {Santiago}, \& {Steinmetz}}]{starhorse}
{Anders}, F., {Khalatyan}, A., {Queiroz}, A.~B.~A., {et~al.} 2022, \aap, 658, A91

\bibitem[{{Andr{\'e}} {et~al.}(2014){Andr{\'e}}, {Di Francesco}, {Ward-Thompson}, {Inutsuka}, {Pudritz}, \& {Pineda}}]{andre2014}
{Andr{\'e}}, P., {Di Francesco}, J., {Ward-Thompson}, D., {et~al.} 2014, in Protostars and Planets VI, ed. H.~{Beuther}, R.~S. {Klessen}, C.~P. {Dullemond}, \& T.~{Henning}, 27--51

\bibitem[{{Andr{\'e}} {et~al.}(2010){Andr{\'e}}, {Men'shchikov}, {Bontemps}, {K{\"o}nyves}, {Motte}, {Schneider}, {Didelon}, {Minier}, {Saraceno}, {Ward-Thompson}, {di Francesco}, {White}, {Molinari}, {Testi}, {Abergel}, {Griffin}, {Henning}, {Royer}, {Mer{\'\i}n}, {Vavrek}, {Attard}, {Arzoumanian}, {Wilson}, {Ade}, {Aussel}, {Baluteau}, {Benedettini}, {Bernard}, {Blommaert}, {Cambr{\'e}sy}, {Cox}, {di Giorgio}, {Hargrave}, {Hennemann}, {Huang}, {Kirk}, {Krause}, {Launhardt}, {Leeks}, {Le Pennec}, {Li}, {Martin}, {Maury}, {Olofsson}, {Omont}, {Peretto}, {Pezzuto}, {Prusti}, {Roussel}, {Russeil}, {Sauvage}, {Sibthorpe}, {Sicilia-Aguilar}, {Spinoglio}, {Waelkens}, {Woodcraft}, \& {Zavagno}}]{andre2010}
{Andr{\'e}}, P., {Men'shchikov}, A., {Bontemps}, S., {et~al.} 2010, \aap, 518, L102

\bibitem[{{Ashby} {et~al.}(2023){Ashby}, {Hora}, {Lakshmipathaiah}, {Vig}, {Sai Subrahmanyam Gorthi}, {Kang}, {Tolls}, {Melnick}, {Werner}, {Crill}, {Masters}, {Pe{\~n}a}, {Lee}, {Kim}, {Lee}, {Yoon}, {Yang}, {Flagey}, \& {Mennesson}}]{ashby2023}
{Ashby}, M. L.~N., {Hora}, J.~L., {Lakshmipathaiah}, K., {et~al.} 2023, \apj, 949, 105

\bibitem[{{Bohlin} {et~al.}(1978){Bohlin}, {Savage}, \& {Drake}}]{bohlin}
{Bohlin}, R.~C., {Savage}, B.~D., \& {Drake}, J.~F. 1978, \apj, 224, 132

\bibitem[{{Carrasco} {et~al.}(2021){Carrasco}, {Weiler}, {Jordi}, {Fabricius}, {De Angeli}, {Evans}, {van Leeuwen}, {Riello}, \& {Montegriffo}}]{carrasco2021}
{Carrasco}, J.~M., {Weiler}, M., {Jordi}, C., {et~al.} 2021, \aap, 652, A86

\bibitem[{{Chambers} {et~al.}(2016){Chambers}, {Magnier}, {Metcalfe}, {Flewelling}, {Huber}, {Waters}, {Denneau}, {Draper}, {Farrow}, {Finkbeiner}, {Holmberg}, {Koppenhoefer}, {Price}, {Rest}, {Saglia}, {Schlafly}, {Smartt}, {Sweeney}, {Wainscoat}, {Burgett}, {Chastel}, {Grav}, {Heasley}, {Hodapp}, {Jedicke}, {Kaiser}, {Kudritzki}, {Luppino}, {Lupton}, {Monet}, {Morgan}, {Onaka}, {Shiao}, {Stubbs}, {Tonry}, {White}, {Ba{\~n}ados}, {Bell}, {Bender}, {Bernard}, {Boegner}, {Boffi}, {Botticella}, {Calamida}, {Casertano}, {Chen}, {Chen}, {Cole}, {Deacon}, {Frenk}, {Fitzsimmons}, {Gezari}, {Gibbs}, {Goessl}, {Goggia}, {Gourgue}, {Goldman}, {Grant}, {Grebel}, {Hambly}, {Hasinger}, {Heavens}, {Heckman}, {Henderson}, {Henning}, {Holman}, {Hopp}, {Ip}, {Isani}, {Jackson}, {Keyes}, {Koekemoer}, {Kotak}, {Le}, {Liska}, {Long}, {Lucey}, {Liu}, {Martin}, {Masci}, {McLean}, {Mindel}, {Misra}, {Morganson}, {Murphy}, {Obaika}, {Narayan}, {Nieto-Santisteban}, {Norberg}, {Peacock}, {Pier}, {Postman}, {Primak}, {Rae}, {Rai},
  {Riess}, {Riffeser}, {Rix}, {R{\"o}ser}, {Russel}, {Rutz}, {Schilbach}, {Schultz}, {Scolnic}, {Strolger}, {Szalay}, {Seitz}, {Small}, {Smith}, {Soderblom}, {Taylor}, {Thomson}, {Taylor}, {Thakar}, {Thiel}, {Thilker}, {Unger}, {Urata}, {Valenti}, {Wagner}, {Walder}, {Walter}, {Watters}, {Werner}, {Wood-Vasey}, \& {Wyse}}]{chambers2016}
{Chambers}, K.~C., {Magnier}, E.~A., {Metcalfe}, N., {et~al.} 2016, arXiv e-prints, arXiv:1612.05560

\bibitem[{{Corradi} {et~al.}(1998){Corradi}, {Aznar}, \& {Mampaso}}]{corradi}
{Corradi}, R. L.~M., {Aznar}, R., \& {Mampaso}, A. 1998, \mnras, 297, 617

\bibitem[{{Crutcher}(2012)}]{Crutcher2012}
{Crutcher}, R.~M. 2012, \araa, 50, 29

\bibitem[{{Cutri} {et~al.}(2003){Cutri}, {Skrutskie}, {van Dyk}, {Beichman}, {Carpenter}, {Chester}, {Cambresy}, {Evans}, {Fowler}, {Gizis}, {Howard}, {Huchra}, {Jarrett}, {Kopan}, {Kirkpatrick}, {Light}, {Marsh}, {McCallon}, {Schneider}, {Stiening}, {Sykes}, {Weinberg}, {Wheaton}, {Wheelock}, \& {Zacarias}}]{Cutri2003}
{Cutri}, R.~M., {Skrutskie}, M.~F., {van Dyk}, S., {et~al.} 2003, {2MASS All Sky Catalog of point sources.}

\bibitem[{{Cutri} {et~al.}(2013){Cutri}, {Wright}, {Conrow}, {Fowler}, {Eisenhardt}, {Grillmair}, {Kirkpatrick}, {Masci}, {McCallon}, {Wheelock}, {Fajardo-Acosta}, {Yan}, {Benford}, {Harbut}, {Jarrett}, {Lake}, {Leisawitz}, {Ressler}, {Stanford}, {Tsai}, {Liu}, {Helou}, {Mainzer}, {Gettings}, {Gonzalez}, {Hoffman}, {Marsh}, {Padgett}, {Skrutskie}, {Beck}, {Papin}, \& {Wittman}}]{cutri2013}
{Cutri}, R.~M., {Wright}, E.~L., {Conrow}, T., {et~al.} 2013, {Explanatory Supplement to the AllWISE Data Release Products}, Explanatory Supplement to the AllWISE Data Release Products, by R. M. Cutri et al.

\bibitem[{{Dzib} {et~al.}(2018){Dzib}, {Loinard}, {Ortiz-Le{\'o}n}, {Rodr{\'\i}guez}, \& {Galli}}]{dzib2018}
{Dzib}, S.~A., {Loinard}, L., {Ortiz-Le{\'o}n}, G.~N., {Rodr{\'\i}guez}, L.~F., \& {Galli}, P. A.~B. 2018, \apj, 867, 151

\bibitem[{{Edenhofer} {et~al.}(2024){Edenhofer}, {Zucker}, {Frank}, {Saydjari}, {Speagle}, {Finkbeiner}, \& {En{\ss}lin}}]{edenhofer2024}
{Edenhofer}, G., {Zucker}, C., {Frank}, P., {et~al.} 2024, \aap, 685, A82

\bibitem[{{Feng} {et~al.}(2024){Feng}, {Smith}, {Hacar}, {Clark}, \& {Seifried}}]{feng2024}
{Feng}, J., {Smith}, R.~J., {Hacar}, A., {Clark}, S.~E., \& {Seifried}, D. 2024, \mnras, 528, 6370

\bibitem[{{Fiege} \& {Pudritz}(2000)}]{fiege2000}
{Fiege}, J.~D. \& {Pudritz}, R.~E. 2000, \mnras, 311, 85

\bibitem[{{Gaia Collaboration} {et~al.}(2021){Gaia Collaboration}, {Brown}, {Vallenari}, {Prusti}, {de Bruijne}, {Babusiaux}, {Biermann}, {Creevey}, {Evans}, {Eyer}, {Hutton}, {Jansen}, {Jordi}, {Klioner}, {Lammers}, {Lindegren}, {Luri}, {Mignard}, {Panem}, {Pourbaix}, {Randich}, {Sartoretti}, {Soubiran}, {Walton}, {Arenou}, {Bailer-Jones}, {Bastian}, {Cropper}, {Drimmel}, {Katz}, {Lattanzi}, {van Leeuwen}, {Bakker}, {Cacciari}, {Casta{\~n}eda}, {De Angeli}, {Ducourant}, {Fabricius}, {Fouesneau}, {Fr{\'e}mat}, {Guerra}, {Guerrier}, {Guiraud}, {Jean-Antoine Piccolo}, {Masana}, {Messineo}, {Mowlavi}, {Nicolas}, {Nienartowicz}, {Pailler}, {Panuzzo}, {Riclet}, {Roux}, {Seabroke}, {Sordo}, {Tanga}, {Th{\'e}venin}, {Gracia-Abril}, {Portell}, {Teyssier}, {Altmann}, {Andrae}, {Bellas-Velidis}, {Benson}, {Berthier}, {Blomme}, {Brugaletta}, {Burgess}, {Busso}, {Carry}, {Cellino}, {Cheek}, {Clementini}, {Damerdji}, {Davidson}, {Delchambre}, {Dell'Oro}, {Fern{\'a}ndez-Hern{\'a}ndez}, {Galluccio}, {Garc{\'\i}a-Lario},
  {Garcia-Reinaldos}, {Gonz{\'a}lez-N{\'u}{\~n}ez}, {Gosset}, {Haigron}, {Halbwachs}, {Hambly}, {Harrison}, {Hatzidimitriou}, {Heiter}, {Hern{\'a}ndez}, {Hestroffer}, {Hodgkin}, {Holl}, {Jan{\ss}en}, {Jevardat de Fombelle}, {Jordan}, {Krone-Martins}, {Lanzafame}, {L{\"o}ffler}, {Lorca}, {Manteiga}, {Marchal}, {Marrese}, {Moitinho}, {Mora}, {Muinonen}, {Osborne}, {Pancino}, {Pauwels}, {Petit}, {Recio-Blanco}, {Richards}, {Riello}, {Rimoldini}, {Robin}, {Roegiers}, {Rybizki}, {Sarro}, {Siopis}, {Smith}, {Sozzetti}, {Ulla}, {Utrilla}, {van Leeuwen}, {van Reeven}, {Abbas}, {Abreu Aramburu}, {Accart}, {Aerts}, {Aguado}, {Ajaj}, {Altavilla}, {{\'A}lvarez}, {{\'A}lvarez Cid-Fuentes}, {Alves}, {Anderson}, {Anglada Varela}, {Antoja}, {Audard}, {Baines}, {Baker}, {Balaguer-N{\'u}{\~n}ez}, {Balbinot}, {Balog}, {Barache}, {Barbato}, {Barros}, {Barstow}, {Bartolom{\'e}}, {Bassilana}, {Bauchet}, {Baudesson-Stella}, {Becciani}, {Bellazzini}, {Bernet}, {Bertone}, {Bianchi}, {Blanco-Cuaresma}, {Boch}, {Bombrun}, {Bossini},
  {Bouquillon}, {Bragaglia}, {Bramante}, {Breedt}, {Bressan}, {Brouillet}, {Bucciarelli}, {Burlacu}, {Busonero}, {Butkevich}, {Buzzi}, {Caffau}, {Cancelliere}, {C{\'a}novas}, {Cantat-Gaudin}, {Carballo}, {Carlucci}, {Carnerero}, {Carrasco}, {Casamiquela}, {Castellani}, {Castro-Ginard}, {Castro Sampol}, {Chaoul}, {Charlot}, {Chemin}, {Chiavassa}, {Cioni}, {Comoretto}, {Cooper}, {Cornez}, {Cowell}, {Crifo}, {Crosta}, {Crowley}, {Dafonte}, {Dapergolas}, {David}, \& {David}}]{Gaia_Collaboration2021}
{Gaia Collaboration}, {Brown}, A.~G.~A., {Vallenari}, A., {et~al.} 2021, \aap, 649, A1

\bibitem[{{Gaia Collaboration} {et~al.}(2016){Gaia Collaboration}, {Prusti}, {de Bruijne}, {Brown}, {Vallenari}, {Babusiaux}, {Bailer-Jones}, {Bastian}, {Biermann}, {Evans}, {Eyer}, {Jansen}, {Jordi}, {Klioner}, {Lammers}, {Lindegren}, {Luri}, {Mignard}, {Milligan}, {Panem}, {Poinsignon}, {Pourbaix}, {Randich}, {Sarri}, {Sartoretti}, {Siddiqui}, {Soubiran}, {Valette}, {van Leeuwen}, {Walton}, {Aerts}, {Arenou}, {Cropper}, {Drimmel}, {H{\o}g}, {Katz}, {Lattanzi}, {O'Mullane}, {Grebel}, {Holland}, {Huc}, {Passot}, {Bramante}, {Cacciari}, {Casta{\~n}eda}, {Chaoul}, {Cheek}, {De Angeli}, {Fabricius}, {Guerra}, {Hern{\'a}ndez}, {Jean-Antoine-Piccolo}, {Masana}, {Messineo}, {Mowlavi}, {Nienartowicz}, {Ord{\'o}{\~n}ez-Blanco}, {Panuzzo}, {Portell}, {Richards}, {Riello}, {Seabroke}, {Tanga}, {Th{\'e}venin}, {Torra}, {Els}, {Gracia-Abril}, {Comoretto}, {Garcia-Reinaldos}, {Lock}, {Mercier}, {Altmann}, {Andrae}, {Astraatmadja}, {Bellas-Velidis}, {Benson}, {Berthier}, {Blomme}, {Busso}, {Carry}, {Cellino}, {Clementini},
  {Cowell}, {Creevey}, {Cuypers}, {Davidson}, {De Ridder}, {de Torres}, {Delchambre}, {Dell'Oro}, {Ducourant}, {Fr{\'e}mat}, {Garc{\'\i}a-Torres}, {Gosset}, {Halbwachs}, {Hambly}, {Harrison}, {Hauser}, {Hestroffer}, {Hodgkin}, {Huckle}, {Hutton}, {Jasniewicz}, {Jordan}, {Kontizas}, {Korn}, {Lanzafame}, {Manteiga}, {Moitinho}, {Muinonen}, {Osinde}, {Pancino}, {Pauwels}, {Petit}, {Recio-Blanco}, {Robin}, {Sarro}, {Siopis}, {Smith}, {Smith}, {Sozzetti}, {Thuillot}, {van Reeven}, {Viala}, {Abbas}, {Abreu Aramburu}, {Accart}, {Aguado}, {Allan}, {Allasia}, {Altavilla}, {{\'A}lvarez}, {Alves}, {Anderson}, {Andrei}, {Anglada Varela}, {Antiche}, {Antoja}, {Ant{\'o}n}, {Arcay}, {Atzei}, {Ayache}, {Bach}, {Baker}, {Balaguer-N{\'u}{\~n}ez}, {Barache}, {Barata}, {Barbier}, {Barblan}, {Baroni}, {Barrado y Navascu{\'e}s}, {Barros}, {Barstow}, {Becciani}, {Bellazzini}, {Bellei}, {Bello Garc{\'\i}a}, {Belokurov}, {Bendjoya}, {Berihuete}, {Bianchi}, {Bienaym{\'e}}, {Billebaud}, {Blagorodnova}, {Blanco-Cuaresma}, {Boch},
  {Bombrun}, {Borrachero}, {Bouquillon}, {Bourda}, {Bouy}, {Bragaglia}, {Breddels}, {Brouillet}, {Br{\"u}semeister}, {Bucciarelli}, {Budnik}, {Burgess}, {Burgon}, {Burlacu}, {Busonero}, {Buzzi}, {Caffau}, {Cambras}, {Campbell}, {Cancelliere}, {Cantat-Gaudin}, {Carlucci}, {Carrasco}, {Castellani}, {Charlot}, {Charnas}, {Charvet}, {Chassat}, {Chiavassa}, {Clotet}, {Cocozza}, {Collins}, {Collins}, \& {Costigan}}]{Gaia_Collaboration2016}
{Gaia Collaboration}, {Prusti}, T., {de Bruijne}, J.~H.~J., {et~al.} 2016, \aap, 595, A1

\bibitem[{{Gaia Collaboration} {et~al.}(2023){Gaia Collaboration}, {Vallenari}, {Brown}, {Prusti}, {de Bruijne}, {Arenou}, {Babusiaux}, {Biermann}, {Creevey}, {Ducourant}, {Evans}, {Eyer}, {Guerra}, {Hutton}, {Jordi}, {Klioner}, {Lammers}, {Lindegren}, {Luri}, {Mignard}, {Panem}, {Pourbaix}, {Randich}, {Sartoretti}, {Soubiran}, {Tanga}, {Walton}, {Bailer-Jones}, {Bastian}, {Drimmel}, {Jansen}, {Katz}, {Lattanzi}, {van Leeuwen}, {Bakker}, {Cacciari}, {Casta{\~n}eda}, {De Angeli}, {Fabricius}, {Fouesneau}, {Fr{\'e}mat}, {Galluccio}, {Guerrier}, {Heiter}, {Masana}, {Messineo}, {Mowlavi}, {Nicolas}, {Nienartowicz}, {Pailler}, {Panuzzo}, {Riclet}, {Roux}, {Seabroke}, {Sordo}, {Th{\'e}venin}, {Gracia-Abril}, {Portell}, {Teyssier}, {Altmann}, {Andrae}, {Audard}, {Bellas-Velidis}, {Benson}, {Berthier}, {Blomme}, {Burgess}, {Busonero}, {Busso}, {C{\'a}novas}, {Carry}, {Cellino}, {Cheek}, {Clementini}, {Damerdji}, {Davidson}, {de Teodoro}, {Nu{\~n}ez Campos}, {Delchambre}, {Dell'Oro}, {Esquej},
  {Fern{\'a}ndez-Hern{\'a}ndez}, {Fraile}, {Garabato}, {Garc{\'\i}a-Lario}, {Gosset}, {Haigron}, {Halbwachs}, {Hambly}, {Harrison}, {Hern{\'a}ndez}, {Hestroffer}, {Hodgkin}, {Holl}, {Jan{\ss}en}, {Jevardat de Fombelle}, {Jordan}, {Krone-Martins}, {Lanzafame}, {L{\"o}ffler}, {Marchal}, {Marrese}, {Moitinho}, {Muinonen}, {Osborne}, {Pancino}, {Pauwels}, {Recio-Blanco}, {Reyl{\'e}}, {Riello}, {Rimoldini}, {Roegiers}, {Rybizki}, {Sarro}, {Siopis}, {Smith}, {Sozzetti}, {Utrilla}, {van Leeuwen}, {Abbas}, {{\'A}brah{\'a}m}, {Abreu Aramburu}, {Aerts}, {Aguado}, {Ajaj}, {Aldea-Montero}, {Altavilla}, {{\'A}lvarez}, {Alves}, {Anders}, {Anderson}, {Anglada Varela}, {Antoja}, {Baines}, {Baker}, {Balaguer-N{\'u}{\~n}ez}, {Balbinot}, {Balog}, {Barache}, {Barbato}, {Barros}, {Barstow}, {Bartolom{\'e}}, {Bassilana}, {Bauchet}, {Becciani}, {Bellazzini}, {Berihuete}, {Bernet}, {Bertone}, {Bianchi}, {Binnenfeld}, {Blanco-Cuaresma}, {Blazere}, {Boch}, {Bombrun}, {Bossini}, {Bouquillon}, {Bragaglia}, {Bramante}, {Breedt},
  {Bressan}, {Brouillet}, {Brugaletta}, {Bucciarelli}, {Burlacu}, {Butkevich}, {Buzzi}, {Caffau}, {Cancelliere}, {Cantat-Gaudin}, {Carballo}, {Carlucci}, {Carnerero}, {Carrasco}, {Casamiquela}, {Castellani}, {Castro-Ginard}, {Chaoul}, {Charlot}, {Chemin}, {Chiaramida}, {Chiavassa}, {Chornay}, {Comoretto}, {Contursi}, {Cooper}, {Cornez}, {Cowell}, {Crifo}, {Cropper}, {Crosta}, {Crowley}, {Dafonte}, {Dapergolas}, {David}, {David}, {de Laverny}, {De Luise}, \& {De March}}]{Gaia_Collaboration2023}
{Gaia Collaboration}, {Vallenari}, A., {Brown}, A.~G.~A., {et~al.} 2023, \aap, 674, A1

\bibitem[{{Ginsburg} \& {Mirocha}(2011)}]{ginsburg11}
{Ginsburg}, A. \& {Mirocha}, J. 2011, {PySpecKit: Python Spectroscopic Toolkit}, Astrophysics Source Code Library, record ascl:1109.001

\bibitem[{{Hacar} {et~al.}(2023){Hacar}, {Clark}, {Heitsch}, {Kainulainen}, {Panopoulou}, {Seifried}, \& {Smith}}]{hacar2023}
{Hacar}, A., {Clark}, S.~E., {Heitsch}, F., {et~al.} 2023, in Astronomical Society of the Pacific Conference Series, Vol. 534, Protostars and Planets VII, ed. S.~{Inutsuka}, Y.~{Aikawa}, T.~{Muto}, K.~{Tomida}, \& M.~{Tamura}, 153

\bibitem[{{Hennebelle} \& {Inutsuka}(2019)}]{Hennebelle2019}
{Hennebelle}, P. \& {Inutsuka}, S.-i. 2019, Frontiers in Astronomy and Space Sciences, 6, 5

\bibitem[{{Jow} {et~al.}(2018){Jow}, {Hill}, {Scott}, {Soler}, {Martin}, {Devlin}, {Fissel}, \& {Poidevin}}]{jow2018}
{Jow}, D.~L., {Hill}, R., {Scott}, D., {et~al.} 2018, \mnras, 474, 1018

\bibitem[{{Kauffmann} {et~al.}(2008){Kauffmann}, {Bertoldi}, {Bourke}, {Evans}, \& {Lee}}]{kauffmann08}
{Kauffmann}, J., {Bertoldi}, F., {Bourke}, T.~L., {Evans}, N.~J., I., \& {Lee}, C.~W. 2008, \aap, 487, 993

\bibitem[{{Koch} \& {Rosolowsky}(2015)}]{koch2015}
{Koch}, E.~W. \& {Rosolowsky}, E.~W. 2015, \mnras, 452, 3435

\bibitem[{{Koenig} \& {Leisawitz}(2014)}]{koening2014}
{Koenig}, X.~P. \& {Leisawitz}, D.~T. 2014, \apj, 791, 131

\bibitem[{{K{\"o}nyves} {et~al.}(2015){K{\"o}nyves}, {Andr{\'e}}, {Men'shchikov}, {Palmeirim}, {Arzoumanian}, {Schneider}, {Roy}, {Didelon}, {Maury}, {Shimajiri}, {Di Francesco}, {Bontemps}, {Peretto}, {Benedettini}, {Bernard}, {Elia}, {Griffin}, {Hill}, {Kirk}, {Ladjelate}, {Marsh}, {Martin}, {Motte}, {Nguy{\^e}n Luong}, {Pezzuto}, {Roussel}, {Rygl}, {Sadavoy}, {Schisano}, {Spinoglio}, {Ward-Thompson}, \& {White}}]{konyv2015}
{K{\"o}nyves}, V., {Andr{\'e}}, P., {Men'shchikov}, A., {et~al.} 2015, \aap, 584, A91

\bibitem[{{Lazarian} \& {Hoang}(2007)}]{Lazarian2007}
{Lazarian}, A. \& {Hoang}, T. 2007, \mnras, 378, 910

\bibitem[{{Li} {et~al.}(2014){Li}, {Goodman}, {Sridharan}, {Houde}, {Li}, {Novak}, \& {Tang}}]{Li2014}
{Li}, H.~B., {Goodman}, A., {Sridharan}, T.~K., {et~al.} 2014, in Protostars and Planets VI, ed. H.~{Beuther}, R.~S. {Klessen}, C.~P. {Dullemond}, \& T.~{Henning}, 101--123

\bibitem[{{Magalhaes} {et~al.}(1996){Magalhaes}, {Rodrigues}, {Margoniner}, {Pereyra}, \& {Heathcote}}]{maglh96}
{Magalhaes}, A.~M., {Rodrigues}, C.~V., {Margoniner}, V.~E., {Pereyra}, A., \& {Heathcote}, S. 1996, in Astronomical Society of the Pacific Conference Series, Vol.~97, Polarimetry of the Interstellar Medium, ed. W.~G. {Roberge} \& D.~C.~B. {Whittet}, 118

\bibitem[{{Nielbock} {et~al.}(2012){Nielbock}, {Launhardt}, {Steinacker}, {Stutz}, {Balog}, {Beuther}, {Bouwman}, {Henning}, {Hily-Blant}, {Kainulainen}, {Krause}, {Linz}, {Lippok}, {Ragan}, {Risacher}, \& {Schmiedeke}}]{Nielbock2012}
{Nielbock}, M., {Launhardt}, R., {Steinacker}, J., {et~al.} 2012, \aap, 547, A11

\bibitem[{{Onken} {et~al.}(2019){Onken}, {Wolf}, {Bessell}, {Chang}, {Da Costa}, {Luvaul}, {Mackey}, {Schmidt}, \& {Shao}}]{Onken2019}
{Onken}, C.~A., {Wolf}, C., {Bessell}, M.~S., {et~al.} 2019, \pasa, 36, e033

\bibitem[{{Ostriker}(1964)}]{Ostriker1964}
{Ostriker}, J. 1964, \apj, 140, 1056

\bibitem[{{Pattle} {et~al.}(2023){Pattle}, {Fissel}, {Tahani}, {Liu}, \& {Ntormousi}}]{Pattle2023}
{Pattle}, K., {Fissel}, L., {Tahani}, M., {Liu}, T., \& {Ntormousi}, E. 2023, in Astronomical Society of the Pacific Conference Series, Vol. 534, Protostars and Planets VII, ed. S.~{Inutsuka}, Y.~{Aikawa}, T.~{Muto}, K.~{Tomida}, \& M.~{Tamura}, 193

\bibitem[{{Pillai} {et~al.}(2020){Pillai}, {Clemens}, {Reissl}, {Myers}, {Kauffmann}, {Lopez-Rodriguez}, {Alves}, {Franco}, {Henshaw}, {Menten}, {Nakamura}, {Seifried}, {Sugitani}, \& {Wiesemeyer}}]{Pillai2020}
{Pillai}, T. G.~S., {Clemens}, D.~P., {Reissl}, S., {et~al.} 2020, Nature Astronomy, 4, 1195

\bibitem[{{Planck Collaboration} {et~al.}(2016){Planck Collaboration}, {Ade}, {Aghanim}, {Alves}, {Arnaud}, {Arzoumanian}, {Ashdown}, {Aumont}, {Baccigalupi}, {Banday}, {Barreiro}, {Bartolo}, {Battaner}, {Benabed}, {Beno{\^\i}t}, {Benoit-L{\'e}vy}, {Bernard}, {Bersanelli}, {Bielewicz}, {Bock}, {Bonavera}, {Bond}, {Borrill}, {Bouchet}, {Boulanger}, {Bracco}, {Burigana}, {Calabrese}, {Cardoso}, {Catalano}, {Chiang}, {Christensen}, {Colombo}, {Combet}, {Couchot}, {Crill}, {Curto}, {Cuttaia}, {Danese}, {Davies}, {Davis}, {de Bernardis}, {de Rosa}, {de Zotti}, {Delabrouille}, {Dickinson}, {Diego}, {Dole}, {Donzelli}, {Dor{\'e}}, {Douspis}, {Ducout}, {Dupac}, {Efstathiou}, {Elsner}, {En{\ss}lin}, {Eriksen}, {Falceta-Gon{\c{c}}alves}, {Falgarone}, {Ferri{\`e}re}, {Finelli}, {Forni}, {Frailis}, {Fraisse}, {Franceschi}, {Frejsel}, {Galeotta}, {Galli}, {Ganga}, {Ghosh}, {Giard}, {Gjerl{\o}w}, {Gonz{\'a}lez-Nuevo}, {G{\'o}rski}, {Gregorio}, {Gruppuso}, {Gudmundsson}, {Guillet}, {Harrison}, {Helou}, {Hennebelle},
  {Henrot-Versill{\'e}}, {Hern{\'a}ndez-Monteagudo}, {Herranz}, {Hildebrandt}, {Hivon}, {Holmes}, {Hornstrup}, {Huffenberger}, {Hurier}, {Jaffe}, {Jaffe}, {Jones}, {Juvela}, {Keih{\"a}nen}, {Keskitalo}, {Kisner}, {Knoche}, {Kunz}, {Kurki-Suonio}, {Lagache}, {Lamarre}, {Lasenby}, {Lattanzi}, {Lawrence}, {Leonardi}, {Levrier}, {Liguori}, {Lilje}, {Linden-V{\o}rnle}, {L{\'o}pez-Caniego}, {Lubin}, {Mac{\'\i}as-P{\'e}rez}, {Maino}, {Mandolesi}, {Mangilli}, {Maris}, {Martin}, {Mart{\'\i}nez-Gonz{\'a}lez}, {Masi}, {Matarrese}, {Melchiorri}, {Mendes}, {Mennella}, {Migliaccio}, {Miville-Desch{\^e}nes}, {Moneti}, {Montier}, {Morgante}, {Mortlock}, {Munshi}, {Murphy}, {Naselsky}, {Nati}, {Netterfield}, {Noviello}, {Novikov}, {Novikov}, {Oppermann}, {Oxborrow}, {Pagano}, {Pajot}, {Paladini}, {Paoletti}, {Pasian}, {Perotto}, {Pettorino}, {Piacentini}, {Piat}, {Pierpaoli}, {Pietrobon}, {Plaszczynski}, {Pointecouteau}, {Polenta}, {Ponthieu}, {Pratt}, {Prunet}, {Puget}, {Rachen}, {Reinecke}, {Remazeilles}, {Renault},
  {Renzi}, {Ristorcelli}, {Rocha}, {Rossetti}, {Roudier}, {Rubi{\~n}o-Mart{\'\i}n}, {Rusholme}, {Sandri}, {Santos}, {Savelainen}, {Savini}, {Scott}, {Soler}, {Stolyarov}, {Sudiwala}, {Sutton}, {Suur-Uski}, {Sygnet}, {Tauber}, {Terenzi}, {Toffolatti}, {Tomasi}, {Tristram}, {Tucci}, {Umana}, {Valenziano}, {Valiviita}, {Van Tent}, {Vielva}, {Villa}, {Wade}, {Wandelt}, {Wehus}, {Ysard}, {Yvon}, \& {Zonca}}]{planck2016}
{Planck Collaboration}, {Ade}, P.~A.~R., {Aghanim}, N., {et~al.} 2016, \aap, 586, A138

\bibitem[{{Ram{\'\i}rez} {et~al.}(2017){Ram{\'\i}rez}, {Magalh{\~a}es}, {Davidson}, {Pereyra}, \& {Rubinho}}]{ramirez2017}
{Ram{\'\i}rez}, E.~A., {Magalh{\~a}es}, A.~M., {Davidson}, Jr., J.~W., {Pereyra}, A., \& {Rubinho}, M. 2017, \pasp, 129, 055001

\bibitem[{{Redaelli} {et~al.}(2019){Redaelli}, {Alves}, {Santos}, \& {Caselli}}]{redaelli2019}
{Redaelli}, E., {Alves}, F.~O., {Santos}, F.~P., \& {Caselli}, P. 2019, \aap, 631, A154

\bibitem[{{Rom{\'a}n-Z{\'u}{\~n}iga} {et~al.}(2010){Rom{\'a}n-Z{\'u}{\~n}iga}, {Alves}, {Lada}, \& {Lombardi}}]{roman10}
{Rom{\'a}n-Z{\'u}{\~n}iga}, C.~G., {Alves}, J.~F., {Lada}, C.~J., \& {Lombardi}, M. 2010, \apj, 725, 2232

\bibitem[{{Roy} {et~al.}(2014){Roy}, {Andr{\'e}}, {Palmeirim}, {Attard}, {K{\"o}nyves}, {Schneider}, {Peretto}, {Men'shchikov}, {Ward-Thompson}, {Kirk}, {Griffin}, {Marsh}, {Abergel}, {Arzoumanian}, {Benedettini}, {Hill}, {Motte}, {Nguyen Luong}, {Pezzuto}, {Rivera-Ingraham}, {Roussel}, {Rygl}, {Spinoglio}, {Stamatellos}, \& {White}}]{roy14}
{Roy}, A., {Andr{\'e}}, P., {Palmeirim}, P., {et~al.} 2014, \aap, 562, A138

\bibitem[{{Seifried} {et~al.}(2019){Seifried}, {Walch}, {Reissl}, \& {Ib{\'a}{\~n}ez-Mej{\'\i}a}}]{Seifried2019}
{Seifried}, D., {Walch}, S., {Reissl}, S., \& {Ib{\'a}{\~n}ez-Mej{\'\i}a}, J.~C. 2019, \mnras, 482, 2697

\bibitem[{{Soler} {et~al.}(2017){Soler}, {Ade}, {Angil{\`e}}, {Ashton}, {Benton}, {Devlin}, {Dober}, {Fissel}, {Fukui}, {Galitzki}, {Gandilo}, {Hennebelle}, {Klein}, {Li}, {Korotkov}, {Martin}, {Matthews}, {Moncelsi}, {Netterfield}, {Novak}, {Pascale}, {Poidevin}, {Santos}, {Savini}, {Scott}, {Shariff}, {Thomas}, {Tucker}, {Tucker}, \& {Ward-Thompson}}]{soler2017}
{Soler}, J.~D., {Ade}, P.~A.~R., {Angil{\`e}}, F.~E., {et~al.} 2017, \aap, 603, A64

\bibitem[{{Tabatabaei} {et~al.}(2024){Tabatabaei}, {Redaelli}, {Galli}, {Caselli}, {Franco}, {Duarte-Cabral}, \& {Padovani}}]{tabatabaei2024}
{Tabatabaei}, F.~S., {Redaelli}, E., {Galli}, D., {et~al.} 2024, \aap, 688, A98

\bibitem[{{Vergely} {et~al.}(2022){Vergely}, {Lallement}, \& {Cox}}]{vergely2022}
{Vergely}, J.~L., {Lallement}, R., \& {Cox}, N.~L.~J. 2022, \aap, 664, A174

\bibitem[{{Wang} {et~al.}(2019){Wang}, {Lai}, {Eswaraiah}, {Pattle}, {Di Francesco}, {Johnstone}, {Koch}, {Liu}, {Tamura}, {Furuya}, {Onaka}, {Ward-Thompson}, {Soam}, {Kim}, {Lee}, {Lee}, {Mairs}, {Arzoumanian}, {Kim}, {Hoang}, {Hwang}, {Liu}, {Berry}, {Bastien}, {Hasegawa}, {Kwon}, {Qiu}, {Andr{\'e}}, {Aso}, {Byun}, {Chen}, {Chen}, {Chen}, {Ching}, {Cho}, {Choi}, {Chrysostomou}, {Chung}, {Coud{\'e}}, {Doi}, {Dowell}, {Drabek-Maunder}, {Duan}, {Eyres}, {Falle}, {Fanciullo}, {Fiege}, {Franzmann}, {Friberg}, {Friesen}, {Fuller}, {Gledhill}, {Graves}, {Greaves}, {Griffin}, {Gu}, {Han}, {Hatchell}, {Hayashi}, {Holland}, {Houde}, {Inoue}, {Inutsuka}, {Iwasaki}, {Jeong}, {Kanamori}, {Kang}, {Kang}, {Kang}, {Kataoka}, {Kawabata}, {Kemper}, {Kim}, {Kim}, {Kim}, {Kim}, {Kirk}, {Kobayashi}, {Konyves}, {Kwon}, {Lacaille}, {Lee}, {Lee}, {Lee}, {Lee}, {Li}, {Li}, {Li}, {Liu}, {Liu}, {Lyo}, {Matsumura}, {Matthews}, {Moriarty-Schieven}, {Nagata}, {Nakamura}, {Nakanishi}, {Ohashi}, {Park}, {Parsons}, {Pascale}, {Peretto},
  {Pon}, {Pyo}, {Qian}, {Rao}, {Rawlings}, {Retter}, {Richer}, {Rigby}, {Robitaille}, {Sadavoy}, {Saito}, {Savini}, {Scaife}, {Seta}, {Shinnaga}, {Tang}, {Tomisaka}, {Tsukamoto}, {van Loo}, {Wang}, {Whitworth}, {Yen}, {Yoo}, {Yuan}, {Yun}, {Zenko}, {Zhang}, {Zhang}, {Zhang}, {Zhou}, \& {Zhu}}]{wang2019}
{Wang}, J.-W., {Lai}, S.-P., {Eswaraiah}, C., {et~al.} 2019, \apj, 876, 42

\bibitem[{{Ward-Thompson} {et~al.}(2007){Ward-Thompson}, {Andr{\'e}}, {Crutcher}, {Johnstone}, {Onishi}, \& {Wilson}}]{ward2007}
{Ward-Thompson}, D., {Andr{\'e}}, P., {Crutcher}, R., {et~al.} 2007, in Protostars and Planets V, ed. B.~{Reipurth}, D.~{Jewitt}, \& K.~{Keil}, 33

\bibitem[{Waskom(2021)}]{seaborn}
Waskom, M.~L. 2021, Journal of Open Source Software, 6, 3021

\bibitem[{{Zhang} {et~al.}(2023){Zhang}, {Green}, \& {Rix}}]{zhang2023}
{Zhang}, X., {Green}, G.~M., \& {Rix}, H.-W. 2023, \mnras, 524, 1855

\bibitem[{{Zucker} \& {Chen}(2018)}]{zucker2018}
{Zucker}, C. \& {Chen}, H. H.-H. 2018, \apj, 864, 152

\end{thebibliography}
%

%----------------
%\begin{thebibliography}{}
%
%  \bibitem[Baker(1966)]{baker} Baker, N. 1966,
%      in Stellar Evolution,
%      ed.\ R. F. Stein,\& A. G. W. Cameron
%      (Plenum, New York) 333
%
%\end{thebibliography}
%---------------

\begin{appendix} %First appendix
\section{Velocity dispersions } \label{dispersion}

\begin{figure}[H]
   \centering
   \includegraphics[width=\hsize]{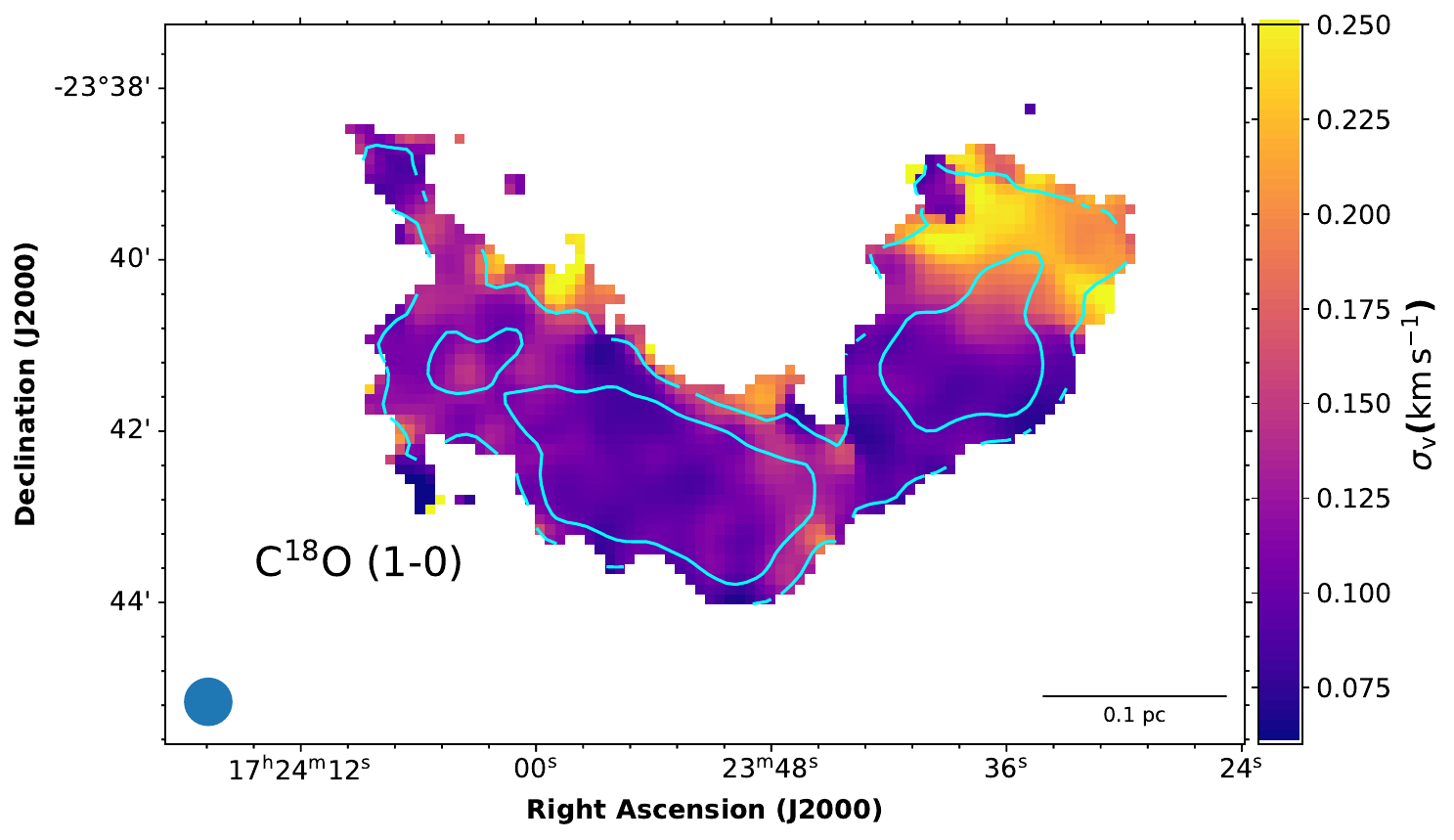}
      \caption{ Velocity dispersion ($\sigma_\mathrm{V}$) map of the \cco\ line. The cyan contours represent the \cco integrated intensity at levels 0.8 and 1.6 K. The beam size is indicated in the bottom-left corner, while the scale bar is displayed in the bottom-right corner.}
         \label{vdisper}
   \end{figure}

The linewidth of the observed \cco\ molecular line was determined by fitting Gaussian profiles to each spectrum in the data cube. Figure~\ref{vdisper} illustrates the derived velocity dispersion map, $\sigma_\mathrm{v}$, highlighting the narrow linewidths of the \cco\ emission, with values typically around 0.1\, \kms to 0.2 \kms. The isothermal sound speed was obtained from:
\begin{equation}
c_\mathrm{s} = \sqrt{\frac{k_\mathrm{B} T}{\mu m_\mathrm{H}}}= 0.2 \;\mathrm{km s^{-1}},
\end{equation}
where $\mu = 2.37$ \citep{kauffmann08} is the mean molecular weight per free particle and $m_\mathrm{H}$ is the mass of hydrogen atom. we used the dust temperature, $T_d = 15$K from \textit{Herschel} map as a proxy for the gas kinetic temperature.

The non-thermal velocity dispersion of the gas can be derived from:
\begin{equation}
\sigma_{\text{nt}} = \sqrt{\sigma_{\text{v}}^2 - \sigma_\mathrm{th}^2}= 0.14 \;\mathrm{km s^{-1}},
\end{equation}
with $\sigma_\mathrm{th}=\sqrt{k_\mathrm{B}T/m_\mathrm{obs}} = 0.06$ \kms and $\sigma_{\text{v}}=0.15$ \kms, where $m_\mathrm{obs}$ is the mass of observed molecule line, \cco.
As mentioned in the main text, the total gas velocity dispersion is calculated using:
\begin{equation} 
\sigma_{\text{tot}} = \sqrt{\sigma_{\text{nt}}^2 + c_\mathrm{s}^2}= 0.24 \;\mathrm{km s^{-1}}. 
\end{equation}

\end{appendix}

\end{document}